%% file: paper_interpolation_ct_unet_efficientnetv2.tex
\documentclass[10pt]{article}

\usepackage[margin=1in]{geometry}
\usepackage[T1]{fontenc}
\usepackage[utf8]{inputenc}
\usepackage{microtype}
\usepackage{amsmath,amssymb}
\usepackage{booktabs}
\usepackage{multirow}
\usepackage{graphicx}
\usepackage{subcaption}
\usepackage{xcolor}
\usepackage[normalem]{ulem}
\usepackage{doi}
\usepackage[numbers,sort&compress]{natbib}
\usepackage{hyperref}
\usepackage{cleveref}
\usepackage{enumitem}
\usepackage{tabularx}
\usepackage{adjustbox}
\usepackage{pdflscape}
\usepackage{tikz}
\usepackage[section]{placeins}
\usetikzlibrary{calc}
\usepackage{fancyhdr}
\usepackage[affil-it]{authblk}

\hypersetup{
  colorlinks=true,
  linkcolor=black,
  citecolor=black,
  urlcolor=darkgray
}

\title{Deep Slice Interpolation for Reducing Through-Plane Anisotropy and Noise in Head CT}
\author[1]{Luis Cortés-Ferre}
\author[1]{Miguel A. Gutiérrez-Naranjo}
\author[2,3]{Marcin Balcerzyk}
\affil[1]{Department of Computer Science and Artificial Intelligence, University of Seville, Avda. Reina Mercedes s/n, 41012 Sevilla, Spain}
\affil[2]{Bioaraba Health Research Institute, New Technologies and Information Systems in Health Research Group, 01009 Vitoria-Gasteiz, Spain}
\affil[3]{IKERBASQUE, Basque Foundation of Science, Plaza Euskadi 5, 48009 Bilbao, Spain}
\date{}

\begin{document}
\maketitle

\begin{abstract}
Head computed tomography (CT) typically uses sub-millimeter in-plane resolution but 2--5\,mm through-plane spacing, creating substantial anisotropy that degrades multiplanar reconstructions, volumetric measurements such as hematoma volume estimation, and downstream algorithms that assume near-isotropic voxels. We present a deep learning system that synthesizes intermediate CT slices from pairs of neighboring axial slices, halving the effective through-plane spacing. The system improves three-dimensional visualization while simultaneously producing inherently denoised outputs, yielding two complementary benefits from a single inference pass.

To build a reliable system, we systematically evaluate pixel-wise losses, namely mean squared error (MSE) and mean absolute error (L1); structural-similarity losses, namely the structural similarity index (SSIM) and its multi-scale variant (MS-SSIM); and hybrid combinations. On a held-out test set, all converged models outperform classical interpolation baselines and pretrained video frame interpolation methods (RIFE, FILM) on all structural measures, with MS-SSIM+L1 offering the strongest balanced profile. We also document training instability in SSIM-family losses and identify partial remedies: the standard numerical fixes eliminate the dominant failure mode but leave residual divergence at smaller batch sizes. All results are reported with patient-level bootstrap confidence intervals and paired statistical tests.

As an illustration, we apply the system to an out-of-distribution head CT series from Hospital Universitario Virgen del Rocío: the model synthesizes intermediate slices and exhibits on the real slices the implicit-denoising signature predicted by our theoretical analysis, supporting in a single external case that interpolation quality and implicit denoising are not confined to the training distribution.
\end{abstract}


\input{sections/intro}

\input{sections/related_work}

\input{sections/methods}
\input{sections/results}
\input{sections/discussion}
\input{sections/conclusion}

\input{sections/acknowledgements}

\input{sections/data_availability}

\input{sections/experiment_scope}

\input{sections/appendix_hemorrhage_tests}
\input{sections/appendix_ssim_k_sensitivity}
\bibliographystyle{unsrtnat}
\bibliography{references}

\end{document}

%% file: sections/intro.tex
\section{Introduction}
\label{sec:intro}
Head computed tomography (CT) volumes are frequently reconstructed with markedly anisotropic voxel spacing, typically exhibiting sub-millimeter in-plane resolution but 2--5\,mm through-plane spacing, a reconstruction convention that improves per-slice signal-to-noise ratio and matches established reading workflows. This anisotropy degrades continuity in multiplanar visualizations, introduces stair-step artifacts in coronal and sagittal views, and adversely affects both visual assessment and downstream pipelines that assume near-isotropic sampling, including volumetric measurements such as hematoma quantification and automated triage systems. Reducing the effective slice spacing through interpolation, if not done during image acquisition or reconstruction, can therefore improve three-dimensional visualization for surgical planning, quantitative analysis, and the reliability of learning-based clinical tools.

We present a deep learning system for head CT slice interpolation that synthesizes an intermediate axial slice $I_{k+1}$ from two neighboring slices $(I_k, I_{k+2})$. The system uses a U-Net with an EfficientNetV2-S encoder, trained on the RSNA 2019 Intracranial Hemorrhage Detection dataset~\cite{ryai2020190211}, which was used in prior work on intracranial hemorrhage detection~\cite{jimaging9020037}. The scope of this work is restricted to non-contrast head CT: the training data, Hounsfield unit (HU) window, and target through-plane spacing are all chosen for this setting, and we make no claim of generalization to other anatomical regions, other modalities, or different acquisition protocols. Cross-anatomy transfer is discussed explicitly as a limitation (\Cref{sec:limitations}) and identified as future work.

The contributions of this work are:
\begin{enumerate}[leftmargin=*]
  \item A \emph{deep learning system for head CT slice interpolation} that outperforms classical baselines and pretrained video frame interpolation models on all structural metrics. Video frame interpolation is a natural transfer source because it addresses the same core problem (synthesizing an intermediate image from two neighboring observations), albeit in the temporal rather than spatial domain; we benchmark RIFE~\cite{RIFE_ECCV2022} and FILM~\cite{reda2022film} to quantify the domain gap.
  \item A \emph{systematic loss function evaluation} under a fixed architecture and training protocol, providing practical guidance on which losses and settings are reliable or unreliable. As part of this analysis, we identify systematic training divergence in SSIM-family losses traceable to numerical conditioning of the SSIM denominator, and provide actionable remedies.
  \item A \emph{finding that regression-based synthesis inherits the denoising property} of conditional-expectation estimation~\cite{bishop2006pattern,hastie2009elements}, as leveraged for image restoration from noisy targets by Lehtinen \emph{et al.}~\cite{noise2noise}, connecting slice interpolation to the low-dose CT denoising literature and delivering interpolation and implicit denoising from a single model.
\end{enumerate}

The remainder of this paper is organized as follows. \Cref{sec:related} reviews through-plane CT super-resolution, video frame interpolation, loss-function comparisons, and low-dose CT denoising. \Cref{sec:methods} describes the dataset, the U-Net with EfficientNetV2-S architecture, the eleven loss configurations evaluated, and the patient-level statistical protocol. \Cref{sec:results} reports held-out test performance with bootstrap confidence intervals and paired Wilcoxon tests, together with a quantitative ROI-level measurement of implicit denoising. \Cref{sec:discussion} analyzes loss--metric alignment, SSIM training instability, implicit denoising as a consequence of regression-based synthesis, an out-of-distribution case study, the effect of pathology on interpolation difficulty, and limitations. \Cref{sec:conclusion} summarizes the findings and outlines directions for future work.

%% file: sections/related_work.tex
\section{Related Work}
\label{sec:related}

\subsection{CT slice interpolation and through-plane super-resolution}
Through-plane resolution enhancement in CT, often termed \emph{super-resolution} (i.e., recovering spatial detail beyond the acquired sampling), has been approached with both classical and learning-based methods.
Spline and polynomial interpolation methods have long been standard in medical image processing~\cite{lehmann816070}, but these methods cannot recover anatomical detail absent from the acquired slices.
Peng \emph{et al.}~\cite{SAINT} proposed SAINT, a spatially aware interpolation network for CT slice synthesis that upsamples volumes along the through-plane axis by an arbitrary integer factor (not limited to the $2\times$ midpoint setting we target), demonstrating that learned models recover structural detail beyond what classical methods produce.
In the broader medical super-resolution literature, Chen \emph{et al.}~\cite{chen-yuhua-xie} applied 3D dense networks to brain MRI super-resolution, and Zhao \emph{et al.}~\cite{SMORE} developed SMORE for through-plane MRI enhancement using self-supervised training on the acquired high-resolution in-plane data.
More recently, Song \emph{et al.}~\cite{Song2024INetII} proposed I$^3$Net, combining inter-slice spatial features with intra-slice frequency-domain learning for CT slice synthesis.
These works establish that through-plane synthesis is a well-studied problem, but systematic comparisons of loss functions within a fixed architecture are rare; most papers report a single loss choice without controlled experiments isolating the effect of individual components. Moreover, these works report aggregate metrics without stratifying evaluation by pathology presence, leaving open whether interpolation quality is maintained in clinically critical abnormal regions. This question is addressed in \Cref{sec:hemorrhage-strat}.

\subsection{Video frame interpolation and cross-domain transfer}
CT slice interpolation is structurally analogous to video frame interpolation (VFI), where an intermediate frame is synthesized from two temporally adjacent frames.
Optical-flow-based VFI methods such as RIFE~\cite{RIFE_ECCV2022} estimate bidirectional intermediate flows with a lightweight IFNet, achieving real-time performance.
FILM~\cite{reda2022film} extends flow-based interpolation to large motions using a multi-scale feature extractor and a frame synthesis network.
Hybrid methods such as AdaCoF~\cite{lee2020adacof} jointly predict spatially adaptive kernel weights and offset vectors, generalizing both kernel-based and flow-based approaches.
While these architectures achieve state-of-the-art results on natural video benchmarks, their direct application to medical imaging is limited: CT slices lack temporal motion, the \emph{displacement} between slices is purely spatial (through-plane anatomy change), and the intensity semantics differ (calibrated HU vs.\ uncalibrated RGB).
Recent work has begun bridging this gap: Gambini \emph{et al.}~\cite{Gambini2024} applied video frame interpolation to MRI and CT volumes, Kim \emph{et al.}~\cite{kim2024dataefficientunsupervisedinterpolationintermediate} proposed UVI-Net for unsupervised 4D medical interpolation, and Sarmad \emph{et al.}~\cite{SARMAD202397} combined transfer learning from pretrained 2D super-resolution networks with a self-supervised consistency loss for 3D volume super-resolution.

\subsection{U-Net and encoder-decoder architectures}
U-Net~\cite{RonnebergerFB15} introduced multiscale feature fusion via skip connections for biomedical image segmentation, and its variants have since become the dominant architecture for medical image segmentation and synthesis.
Replacing the U-Net encoder with pretrained classification backbones, a strategy popularized by open-source toolkits~\cite{Iakubovskii:2019}, reuses the visual feature detectors (edges, textures, shapes) that convolutional filters learn during ImageNet pretraining, enabling effective training on medical tasks with limited labelled data.
EfficientNetV2~\cite{Tan2021EfficientNetV2SM} offers a favorable parameter-accuracy trade-off through fused MBConv blocks, making it an efficient encoder choice.
For image synthesis specifically, Isola \emph{et al.}~\cite{IsolaZZE16} demonstrated that U-Net generators with skip connections outperform encoder-decoder alternatives in paired image-to-image translation, a finding relevant to medical imaging contexts where spatial correspondence between input and output is preserved.

\subsection{Loss functions, training stability, and numerical conditioning}
The choice of training loss significantly affects reconstruction quality, yet systematic comparisons are sparse.
Zhao \emph{et al.}~\cite{Zhao2017LossFF} compared L1, L2, SSIM, and MS-SSIM losses across image restoration tasks including super-resolution, finding that a combined loss formed as a weighted sum of MS-SSIM and L1 achieves the best scores on both distortion and perceptual quality metrics.
SSIM~\cite{wang1284395} and its multiscale extension MS-SSIM~\cite{wang1292216} optimize windowed structural similarity, emphasizing luminance, contrast, and structure rather than pointwise error.
However, SSIM-based losses can exhibit numerical instability in practice: the small constants that stabilize the SSIM denominator must be set carefully to avoid division-by-near-zero in low-variance image regions.
Kastryulin \emph{et al.}~\cite{kastryulin2022pytorchimagequalitymetrics} surveyed differentiable image quality metrics and found that even correct SSIM implementations produce inconsistent correlation values across libraries.
More broadly, deep learning papers overwhelmingly report successful runs, yet training instability is common in practice.
Reinke \emph{et al.}~\cite{Reinke2024} highlighted that metric selection and reporting practices vary widely in medical imaging and advocated for standardized evaluation frameworks.
Training divergence patterns, i.e., configurations that fail to converge or produce degenerate outputs, are rarely analyzed systematically, yet they carry direct engineering value for practitioners adopting these losses.

\subsection{Low-dose CT denoising and implicit noise suppression}
CT noise arises from photon counting statistics and electronic interference, is spatially correlated and reconstruction-kernel-dependent, and degrades diagnostic performance, particularly at reduced radiation doses and high axial resolution; recent reviews survey the deep-learning-based methods that address this problem~\cite{CHEN2024111355}.

Modern CT scanners compensate for the increased noise associated with low-dose acquisition through a combination of hardware and algorithmic strategies that act across the entire imaging chain. Advances in detector efficiency, bow-tie filtration, and automatic tube-current modulation improve photon utilization and reduce quantum noise at the source, while projection-space preprocessing suppresses stochastic fluctuations before reconstruction~\cite{manduca2009}. Contemporary reconstruction methods, particularly iterative reconstruction~\cite{Klink2014-sm} and, more recently, deep-learning-based reconstruction~\cite{Chen2017-ml}, further mitigate noise by incorporating accurate system models, statistical priors, and learned anatomical features, enabling substantial noise reduction without compromising spatial resolution or texture fidelity. These techniques act during acquisition and reconstruction; the regression-based denoising described below, and inherited by our interpolation model, is complementary: it operates on already-reconstructed volumes where upstream options are no longer available.

Deep learning approaches to low-dose CT reconstruction, notably encoder--decoder networks trained with MSE loss, have demonstrated simultaneous noise suppression, structural preservation, and improved lesion detection in a clinical reader study~\cite{Chen2017-ml}.
The underlying statistical mechanism is the classical regression identity that the predictor minimizing the mean squared error equals the conditional expectation of the target~\cite{bishop2006pattern,hastie2009elements}; when the noise on the target averages to zero given the inputs (a \emph{zero-conditional-mean} assumption, i.e., the inputs do not systematically bias the noise up or down), this conditional expectation equals the underlying signal, so regression-trained networks act as implicit denoisers even when the training targets themselves are noisy; in other words, the model tends to recover the underlying signal rather than reproduce the noise.

Lehtinen \emph{et al.}~\cite{noise2noise} leveraged this identity to show that a clean training target is not required. They trained image-restoration networks on pairs consisting of \emph{two independent noisy observations of the same underlying scene}, using one as the input and the other as the target under a standard regression loss. Because the two noise terms are independent and satisfy the zero-conditional-mean assumption, the regression identity still makes the clean signal the minimizer of the loss in expectation, so the trained network recovers that signal even though no clean image was ever used as a target. This is the central finding of Noise2Noise: supervised denoising is possible without clean targets. In our setup, we do not have two noisy measurements of the same slice, but we do have three distinct adjacent slices: the inputs $(I_k, I_{k+2})$ and the target $I_{k+1}$ each carry their own noise, and the underlying signal itself differs across them because anatomy varies along the through-plane axis. What transfers is therefore not the Noise2Noise construction but the underlying conditional-expectation identity itself, and our interpolation model, trained with regression losses on noisy CT slices, inherits the same implicit-denoising behavior (see \Cref{sec:noise} for a precise statement and derivation).

%% file: sections/methods.tex
\section{Materials and Methods}
\label{sec:methods}

\subsection{Dataset and preprocessing}
\label{sec:data}
We use the RSNA Intracranial Hemorrhage Detection dataset~\cite{ryai2020190211}, which provides slice-level multi-label annotations for five hemorrhage subtypes (epidural, subdural, subarachnoid, intraparenchymal, intraventricular) plus an \emph{any}-hemorrhage flag. These categories are RSNA challenge labels rather than ICD-10/ICD-11 hemorrhage subtypes. The dataset is distributed through the Kaggle competition platform under the competition's data-use agreement; our training code, model configurations, and evaluation scripts are released separately (see \Cref{sec:appendix_repro}). The RSNA dataset is multi-institutional with heterogeneous acquisition protocols; reported inter-slice spacing typically falls in the 2.5--5\,mm range with sub-millimeter in-plane resolution. Synthesizing one intermediate slice between each pair of acquired slices halves the effective through-plane spacing from ${\sim}5$\,mm to ${\sim}2.5$\,mm.

DICOM slices are windowed with center $44$\,HU and width $128$\,HU. We clamp the upper bound at $107$\,HU so that the windowed interval covers $128$ integer HU steps and discretizes uniformly to the $256$ levels of $8$-bit PNG storage (two levels per HU); the effective HU window is therefore $[-20\,\text{HU}, 107\,\text{HU}]$. The windowed slices are converted to grayscale PNG and normalized to $[0,1]$, so that $-20\,\text{HU}$ maps to $0$ and $107\,\text{HU}$ to $1$. All training, evaluation, and interpolation operate on these $[0,1]$-normalized intensities; for clarity we report intensity ranges in HU units throughout the paper. Values outside $[-20\,\text{HU}, 107\,\text{HU}]$ are saturated at preprocessing, and the model is neither trained nor calibrated for out-of-window content.

The prior hemorrhage detection work~\cite{jimaging9020037} used three HU windows stacked as RGB channels. For the single-channel interpolation task, we selected this window to cover brain parenchyma (${\sim}20$--$45$\,HU) and acute hemorrhage (${\sim}60$--$80$\,HU); it is wider than the standard brain window (center $40$\,HU, width $80$\,HU, range $[0\,\text{HU}, 80\,\text{HU}]$) and leaves headroom on both sides of the clinically relevant intensities. Air (below $-20$\,HU) and bone (above $107$\,HU) still fall outside the window.

We use a standard patient-level split into training, validation, and test sets: 30 patients are held out as the test set, and the remaining 18{,}870 patients are split 80/20 into training (15{,}096 patients, 570{,}562 triplets) and validation (3{,}774 patients, 141{,}747 triplets). $9.3\%$ of RSNA patients ($1{,}755$ of $18{,}900$) are associated with multiple CT studies; we treat each study as an independent sequence when constructing triplets, so no triplet spans two studies. The patient-level split assigns all studies of a given patient to a single partition, and the 30 held-out test patients are additionally restricted to those with exactly one study so that patient-level paired comparisons are unambiguous.

The 30 test patients were selected to enable the patient-level paired tests reported in \Cref{sec:paired}. From the subset of patients with exactly one CT study, patients were selected to enrich the test set for pathology while guaranteeing representation of every hemorrhage subtype: for each of the 5 subtypes, 5 patients containing that subtype were randomly selected (without replacement across subtypes) and removed from the candidate pool; 5 additional patients were then drawn uniformly at random from the remaining pool, which contained both hemorrhage and normal studies. This yielded 27 patients carrying at least one hemorrhage subtype and 3 purely normal patients, totaling 968 interpolation triplets. The pathology enrichment is intentional: the clinically motivated question is whether interpolation quality holds on abnormal anatomy, where structural fidelity matters most; the 3 normal patients (572 slice-level triplets) provide a sanity check that performance does not degrade on normal cases. At the slice level, 396 triplets (41\%) contain hemorrhage and 572 (59\%) do not; this hemorrhage prevalence is substantially higher than in unselected head-CT cohorts. The test-set patient IDs are released as a frozen artefact. Hemorrhage labels are multi-label (190 of 396 triplets carry more than one subtype), so ``any-of'' counts exceed the total: subdural ($n=189$), subarachnoid ($n=184$), intraparenchymal ($n=132$), intraventricular ($n=114$), and epidural ($n=44$); these per-subtype counts support the hemorrhage-stratified analysis of \Cref{sec:hemorrhage-strat}.

\subsection{Task definition}
\label{sec:task}
Given consecutive triplets $(I_k, I_{k+1}, I_{k+2})$, input is $X=[I_k, I_{k+2}]$ (2 channels), target is $Y=I_{k+1}$, and the model's prediction is denoted $\hat Y$. This formulation assumes that the target slice is equidistant from the two input slices. The RSNA dataset is largely acquired at a single fixed inter-slice spacing per patient, so this assumption holds for the majority of cases; a small number of test patients contain variable-spacing series, for which the equidistance assumption is approximate.

\subsection{Model and augmentation}
\label{sec:model}

\input{figures/architecture}

We hold the network architecture fixed across all experiments in order to isolate the impact of training design choices. Accordingly, we do not claim universality of the reported findings across different architectural designs.
The architecture is a U-Net~\cite{RonnebergerFB15} with an EfficientNetV2-S encoder~\cite{Tan2021EfficientNetV2SM} (\Cref{fig:architecture}); implementation libraries and versions are listed in \Cref{sec:appendix_repro}. The encoder produces features at four skip-connection levels (channel dimensions $24, 48, 64, 160$ at strides $2, 4, 8, 16$) plus a $256$-channel bottleneck at stride $32$. The bottleneck feeds the decoder, and the four skip-connection levels are concatenated into the corresponding decoder stages. The decoder consists of five upsampling blocks with output channels $(256, 128, 64, 32, 16)$; each block doubles the spatial resolution, concatenates the upsampled feature map with its encoder skip (the final block, operating at full resolution, has no skip), and applies two $3\times3$ Conv--BatchNorm--ReLU sub-blocks. A final $3\times3$ convolution maps the 16-channel feature map to a single-channel output, with no output activation applied: the task is regression, and the target range $[0,1]$ is controlled by the normalized data and the bounded losses of \Cref{sec:losses}. The encoder is initialized from ImageNet-pretrained weights; the decoder and segmentation head are initialized from scratch: decoder convolutions use Kaiming-uniform weight initialization for ReLU~\cite{he2015delving} (fan-in mode, zero bias), batch normalization layers are set to (weight $=1$, bias $=0$), and the final $3\times3$ head uses Xavier-uniform weight initialization~\cite{glorot2010understanding} with zero bias. Because ImageNet pretraining uses 3-channel RGB input while our task has 2-channel input $(I_k, I_{k+2})$, the first convolutional layer (stem) is adapted by keeping the first two of the three pretrained RGB filters (dropping the blue channel) and rescaling them by $3/2$ to approximately preserve the activation magnitude at the stem output. \Cref{tab:model_complexity} summarizes the model's computational profile.

\begin{table}[t]
  \centering
  \caption{Model complexity (U-Net + EfficientNetV2-S). Training and test-time inference both operate on $256 \times 256$ patches; each test prediction is reassembled from 13 patches (9 overlapping center crops + 4 corner patches) into a $512 \times 512$ composite (see \Cref{sec:model}). GFLOPs are reported as $2\times$ multiply--accumulate operations (the convention used by fvcore and ptflops).}
  \label{tab:model_complexity}
  \small
  \begin{tabular}{lr}
    \toprule
    Property & Value \\
    \midrule
    Total parameters & 22.1\,M \\
    \quad Encoder (pretrained) & 19.8\,M \\
    \quad Decoder (random init) & 2.2\,M \\
    \quad Segmentation head ($3\!\times\!3$ conv) & $<$0.1\,M \\
    GFLOPs per $256 \times 256$ patch & 12.4 \\
    GFLOPs per reconstructed slice (13 patches) & 161 \\
    \bottomrule
  \end{tabular}
\end{table}

\Cref{fig:augmentation} shows the patch geometry shared by training-time sampling and test-time inference. Each training sample draws one $256 \times 256$ view from a discrete pool: the nine overlapping center crops of \Cref{fig:augmentation}(a), plus a full-slice Lanczos~\cite{duchon1979lanczos} downsample to $256 \times 256$ (not shown in the figure). One of these ten options is drawn per training sample with weighted random sampling: the central crop carries twice the weight of each off-center crop, and the Lanczos option carries half. Normalizing the weights to sum to $1$ yields selection probabilities of $\approx\!0.19$ for the central crop, $\approx\!0.095$ for each of the eight off-center crops, and $\approx\!0.048$ for the Lanczos downsample. After the spatial pick we apply an independent $50\%$ horizontal flip, $50\%$ vertical flip, and a $50\%$ chance of rotation uniformly drawn from $[-15^\circ, +15^\circ]$; all three transforms are applied identically to $I_k$, $I_{k+1}$, and $I_{k+2}$ to preserve triplet alignment. At test time, each $512 \times 512$ slice is reconstructed from 13 $256 \times 256$ patches: the same nine overlapping center crops (\Cref{fig:augmentation}(a)) are averaged over their overlaps to produce the central $384 \times 384$ band, and four non-overlapping corner patches $C_1$--$C_4$ (\Cref{fig:augmentation}(b)) tile the full slice; the averaged nine-crop band is then overlaid on top, so only the outer $64$-pixel strip of each corner patch contributes to the final composite (\Cref{fig:augmentation}(c)). The composite is compared against the $512 \times 512$ target, preserving direct comparability with the classical and VFI baselines which produce full-resolution predictions natively. Validation drives early stopping and best-epoch checkpoint selection; it evaluates the model on each of the nine $256 \times 256$ center crops and averages the resulting metrics.

\input{figures/augmentation}

\subsection{Loss families and numerical stability}
\label{sec:losses}

\noindent We evaluate four single-term losses (two pixel-wise and two structural) and two combined-loss families (SSIM+L1 and MS-SSIM+L1) whose mixing weight $\alpha$ is swept across configurations. All losses operate on predictions $\hat Y$ and targets $Y$ normalized to $[0,1]$ and indexed over $N$ pixels.

\paragraph{Pixel-wise losses.}
The two pixel-wise losses are the mean squared error and the mean absolute error,
\begin{equation}
\mathcal{L}_{\mathrm{MSE}}(\hat Y, Y) \;=\; \frac{1}{N}\sum_{i=1}^{N}\bigl(\hat y_i - y_i\bigr)^{2},
\qquad
\mathcal{L}_{\mathrm{L1}}(\hat Y, Y) \;=\; \frac{1}{N}\sum_{i=1}^{N}\bigl\lvert \hat y_i - y_i \bigr\rvert.
\label{eq:pixel_losses}
\end{equation}
Both are standard regression losses whose minimizers are well-known conditional statistics: $\mathcal{L}_{\mathrm{MSE}}$ has the conditional mean $\mathbb{E}[Y \mid X]$ as its minimizer, and $\mathcal{L}_{\mathrm{L1}}$ the conditional median~\cite{bishop2006pattern,hastie2009elements}; this property is what underpins the implicit-denoising argument developed in \Cref{sec:noise}. The theoretical argument applies to these pixel-wise losses specifically; SSIM-based losses are not conditional-mean or -median estimators and so fall outside it, but are nonetheless measured to denoise empirically at reduced strength (\Cref{sec:noise-results}).

\paragraph{Structural losses.}
The single-scale structural similarity index (SSIM)~\cite{wang1284395} is computed on sliding $11\times 11$ Gaussian-weighted windows with $\sigma = 1.5$. For aligned windows $x, y$ with local means $\mu_x, \mu_y$, variances $\sigma_x^{2}, \sigma_y^{2}$, and covariance $\sigma_{xy}$,
\begin{equation}
\mathrm{SSIM}(x,y) \;=\; \frac{\bigl(2\mu_x \mu_y + C_1\bigr)\bigl(2\sigma_{xy} + C_2\bigr)}{\bigl(\mu_x^{2} + \mu_y^{2} + C_1\bigr)\bigl(\sigma_x^{2} + \sigma_y^{2} + C_2\bigr)},
\qquad
C_k \;=\; (K_k L)^{2},
\label{eq:ssim}
\end{equation}
where $L = 1$ is the dynamic range of our $[0,1]$-normalized intensities (\Cref{sec:data}) and $(K_1, K_2)$ are small dimensionless scalars whose only role is to keep the denominator bounded away from zero in low-variance regions. Wang \emph{et al.}~\cite{wang1284395} propose the defaults $(K_1, K_2) = (0.01, 0.03)$, which we retain for the \emph{evaluation metric}, so that the reported SSIM numbers are comparable to prior literature and preserve the discriminating power characterised in \Cref{sec:appendix_ssim_k}, but modify for training as a numerical-conditioning knob (see below). The image-level index $\overline{\mathrm{SSIM}}(\hat Y, Y)$ is the mean of \Cref{eq:ssim} over all windows, and the loss is $\mathcal{L}_{\mathrm{SSIM}} = 1 - \overline{\mathrm{SSIM}}(\hat Y, Y)$.

Multi-scale SSIM~\cite{wang1292216} decomposes \Cref{eq:ssim} as $\mathrm{SSIM} = l \cdot c \cdot s$ (luminance, contrast, and structure) and evaluates these components at $M=5$ dyadically downsampled resolutions, indexed $j=1$ (finest) to $j=M$ (coarsest). Contrast $c_j$ and structure $s_j$ are aggregated across all scales, while luminance is retained only at the coarsest scale, $l_M$, since it is a low-frequency property that does not benefit from multi-scale repetition:
\begin{equation}
\mathrm{MS\text{-}SSIM}(\hat Y, Y) \;=\; \bigl[l_M(\hat Y, Y)\bigr]^{\alpha_M} \prod_{j=1}^{M} \bigl[c_j(\hat Y, Y)\bigr]^{\beta_j} \bigl[s_j(\hat Y, Y)\bigr]^{\gamma_j},
\label{eq:msssim}
\end{equation}
with exponents fixed to the calibrated weights of Wang\footnote{We follow the standard convention $\beta_j = \gamma_j = w_j$ for $j = 1,\dots,M$ with $\alpha_M = \beta_M$, and $\sum_j w_j = 1$, using $w = (0.0448, 0.2856, 0.3001, 0.2363, 0.1333)$.}~\emph{et al.}~\cite{wang1292216}. The training loss is $\mathcal{L}_{\mathrm{MS\text{-}SSIM}} = 1 - \mathrm{MS\text{-}SSIM}(\hat Y, Y)$.

\paragraph{Combined losses.}
Combined losses pair a structural term with a pixel-wise term, following Zhao \emph{et al.}~\cite{Zhao2017LossFF}, to combine the low-frequency fidelity of pixel-wise losses with the local structural sensitivity of SSIM. The two combined-loss families are
\begin{align}
\mathcal{L}_{\mathrm{SSIM+L1}}(\alpha) &\;=\; \alpha\,\mathcal{L}_{\mathrm{SSIM}} \;+\; (1-\alpha)\,\mathcal{L}_{\mathrm{L1}}, \label{eq:ssim_l1} \\
\mathcal{L}_{\mathrm{MS\text{-}SSIM+L1}}(\alpha) &\;=\; \alpha\,\mathcal{L}_{\mathrm{MS\text{-}SSIM}} \;+\; (1-\alpha)\,\mathcal{L}_{\mathrm{L1}}, \label{eq:msssim_l1}
\end{align}
with $\alpha \in \{0.2,\, 0.3,\, 0.5,\, 0.8\}$ swept across configurations; the primary reported model and the reference for all paired tests is $\mathcal{L}_{\mathrm{MS\text{-}SSIM+L1}}$ at $\alpha = 0.5$.

\paragraph{Numerical conditioning of SSIM-based training losses.}
\label{par:ssim_stability}
SSIM and MS-SSIM are rational functions whose denominator $(\mu_x^{2} + \mu_y^{2} + C_1)(\sigma_x^{2} + \sigma_y^{2} + C_2)$ can collapse towards its $C$-floors in low-texture regions, primarily through the contrast/structure factor $\sigma_x^{2} + \sigma_y^{2} + C_2$ in uniform brain parenchyma and large CSF pools, and through both factors in background air (which clips to $0$ under our normalization). With $L = 1$ the Wang \emph{et al.}\ defaults evaluate to $C_1 = 10^{-4}$ and $C_2 = 9\!\times\!10^{-4}$. On a random sample of 5000 training slices under the brain window, we compute the $11\!\times\!11$ Gaussian-weighted local variance (the same window operator that appears inside SSIM, with $\sigma = 1.5$) at every pixel of every slice. Restricted to windows whose local mean intensity exceeds $0.15$ (a threshold that corresponds to approximately $-1$\,HU under the $[-20\,\text{HU}, 107\,\text{HU}]$ brain window, chosen to exclude background air while retaining soft tissue, CSF, parenchyma, and bone), the median local variance is $1.0\!\times\!10^{-3}$, numerically indistinguishable from the default $C_2 = 9\!\times\!10^{-4}$. In those windows the denominator $\sigma_x^{2} + \sigma_y^{2} + C_2$ is dominated by the stability floor rather than by signal, and the resulting near-zero-denominator regime produces large, oscillatory gradients and overflow to NaN during early training. We therefore combine two interventions, each addressing a distinct failure mode of the raw formulation:
\begin{enumerate}[leftmargin=*,itemsep=1pt]
  \item \textbf{Larger $K_2$.} We set $K_1 = 0.01$ and $K_2 = 0.4$ for all SSIM-based training losses, raising $C_2$ from $9\!\times\!10^{-4}$ to $0.16$ (roughly two orders of magnitude). On the same 5000-slice sample, $C_2 = 0.16$ exceeds $99.0\%$ of all window variances and $96.5\%$ of those above the background-air threshold defined above, so the denominator-stability floor of \Cref{eq:ssim} dominates the natural within-window variance in all but the highest-contrast edges. $K_1 = 0.01$ is retained because the luminance-term denominator $\mu_x^{2} + \mu_y^{2} + C_1$ is not the limiting factor in our data. The choice is restricted to the training loss; test-set SSIM and MS-SSIM remain reported at the Wang \emph{et al.}\ defaults for comparability with prior work. \Cref{sec:appendix_ssim_k} quantifies the evaluation-metric sensitivity to $K_2$: on the same test predictions, reporting at $K_2 = 0.4$ rather than $0.03$ uniformly inflates mean SSIM by $\approx 0.09$ but compresses the inter-model spread by a factor of $1.57$, so the Wang \emph{et al.}\ choice at evaluation is both the literature-comparable and the more discriminating setting.
  \item \textbf{FP32 autocast exclusion.} SSIM's variance and covariance terms are second-order statistics computed via the identity $\sigma^{2} = \mathbb{E}[X^{2}] - (\mathbb{E}[X])^{2}$, i.e., the mean of the squared pixel values in the window minus the squared window mean. In low-texture windows these two quantities are close, so their difference is the small tail left after many matching leading digits. This subtraction of near-equal numbers is catastrophic in half precision (\texttt{float16}, the IEEE 16-bit floating-point format used by default in mixed-precision training): most of the matching digits are lost, and the result can emerge with essentially random sign. Mathematically $\sigma^{2}$ is non-negative; computed through this identity in \texttt{float16}, it can be returned slightly negative. This contaminates two operations in implementations that follow this algebraic form: the $\sigma_x^{2} + \sigma_y^{2} + C_2$ denominator of \Cref{eq:ssim} can flip sign or pass through zero, and the MS-SSIM aggregate $\prod_{j} (c_j s_j)^{w_j}$ raises possibly-negative per-scale factors to fractional exponents $w_j$, yielding NaN; the resulting NaN gradient then propagates through backpropagation and terminates the run. We therefore compute every SSIM and MS-SSIM forward pass in full \texttt{float32} precision\footnote{The SSIM subgraph runs with the autocast context disabled and its inputs cast to \texttt{float32}; the convolutional backbone remains in mixed precision, the PyTorch~AMP default that keeps most operators in half precision while automatically promoting numerically sensitive reductions to \texttt{float32}.}, while the convolutional backbone remains in mixed precision for throughput.
\end{enumerate}
Interventions~1--2 eliminated the most common failure mode (within-epoch NaN at initialization) but did not eliminate training-time divergence (\Cref{par:divergence-definition}) entirely: a substantial fraction of SSIM-family configurations at smaller batch sizes and at particular learning rates still diverged. The full divergence structure and its dependence on learning rate and batch size are analyzed in \Cref{sec:ssim-fragility}, and the per-family, per-batch-size pass/fail accounting is given in Appendix~\ref{sec:appendix_registry}.

\paragraph{Hyperparameter sweep.}
Each loss family was trained across learning rates in the range $[10^{-4},\, 3\!\times\!10^{-3}]$ at the default batch size of 96. SSIM-family configurations additionally include smaller batch sizes where indicated, and three non-SSIM runs (two L1, one MSE) are repeated at batch size 64 as a stability check; the per-family, per-batch-size coverage is tabulated in \Cref{tab:divergence_accounting}. The mixing weight $\alpha \in \{0.2, 0.3, 0.5, 0.8\}$ is varied in combined-loss configurations as in Equations~(\ref{eq:ssim_l1})--(\ref{eq:msssim_l1}). Five MS-SSIM+L1 recovery-attempt fine-tunes, additional to this systematic sweep, start from diverged MS-SSIM+L1 checkpoints at learning rate (lr) $\in \{5\!\times\!10^{-5},\, 10^{-4},\, 3\!\times\!10^{-4}\}$, batch size (bs) $= 64$, and a 300-epoch budget (constant LR for three of the five); all five diverged and are included in the divergence accounting of Appendix~\ref{sec:appendix_registry} but do not enter any reported metric. All remaining hyperparameters are fixed as described in \Cref{sec:training}.

\subsection{Optimization and training protocol}
\label{sec:training}
All experiments use AdamW~\cite{loshchilov2018decoupled} with default $(\beta_1, \beta_2, \varepsilon)$ and weight decay $0.01$. The learning rate follows a linear warmup over the first 5 epochs from $0$ to the configured peak and then cosine-anneals to $\eta_{\min} = 10^{-6}$ over the remaining 495 epochs of the 500-epoch budget; gradients are clipped at $\lVert\cdot\rVert_2 \leq 1$. Training is mixed-precision, with the SSIM and MS-SSIM forward passes kept in \texttt{float32} as described in intervention~2 of \Cref{sec:losses}. Early stopping monitors the validation loss with patience~$15$ and minimum delta~$10^{-4}$; the monitored quantity is loss-family-specific by construction, so each run is halted against its own validation trajectory rather than a shared surrogate. The train/validation split is patient-wise (no slices from a validation patient appear in training). \Cref{tab:hyperparams} lists the shared hyperparameters; environment, hardware, seeding, and determinism settings are reported in \Cref{sec:appendix_repro}.

\begin{table}[t]
  \centering
  \caption{Shared training hyperparameters across all experiments.}
  \label{tab:hyperparams}
  \small
  \begin{tabular}{ll}
    \toprule
    Parameter & Value \\
    \midrule
    Optimizer & AdamW, $(\beta_1, \beta_2, \varepsilon) = (0.9, 0.999, 10^{-8})$, weight decay $= 0.01$ \\
    Scheduler & Linear warmup (5 epochs, $0 \to \text{peak}$), then cosine to $\eta_{\min} = 10^{-6}$ \\
    Gradient clipping & $\lVert\cdot\rVert_2 \leq 1$ \\
    Max epochs & 500 (early stopping triggers first) \\
    Early stopping & monitor: validation loss (min); patience $= 15$; min.\ delta $= 10^{-4}$ \\
    Default batch size & 96 (varied: 32, 64, 96) \\
    SSIM window & $11 \times 11$ (Wang \emph{et al.} default) \\
    MS-SSIM scales & 5 (Wang \emph{et al.} default) \\
    Train / validation split & 80\% / 20\% of non-test patients, patient-wise \\
    Test set & 30 patients (subtype-stratified selection) \\
    Hardware & 1$\times$ NVIDIA GeForce RTX 3080 Ti (12~GB) \\
    \bottomrule
  \end{tabular}
\end{table}

\subsection{Evaluation metrics}
\label{sec:metrics}
We report five complementary metrics: the structural similarity index (SSIM) and its multi-scale variant (MS-SSIM), mean absolute error (MAE), gradient MAE, and peak signal-to-noise ratio (PSNR). Higher is better for SSIM, MS-SSIM, and PSNR; lower is better for MAE and gradient MAE. All metrics are computed on the reconstructed $512 \times 512$ prediction and target slices (\Cref{fig:augmentation}(c)) in the same $[0,1]$-normalized representation used for training, i.e.\ after the center-$44$\,HU / width-$128$\,HU window of \Cref{sec:data}; no further windowing or rescaling is applied at evaluation time. The concrete definitions are as follows.

\begin{itemize}[leftmargin=*, itemsep=1pt]
  \item \textbf{SSIM, MS-SSIM.} As defined by Equations~(\ref{eq:ssim})--(\ref{eq:msssim}) with $11\times 11$ Gaussian windows, $M=5$ scales, the calibrated weights of Wang \emph{et al.}~\cite{wang1292216}, $w = (0.0448, 0.2856, 0.3001, 0.2363, 0.1333)$, and the Wang \emph{et al.}~\cite{wang1284395} defaults $(K_1, K_2) = (0.01,\, 0.03)$ restored for evaluation, in contrast to the training-only choice $K_2 = 0.4$ motivated in intervention~1 of \Cref{sec:losses}.
  \item \textbf{MAE.} Pixel-wise mean absolute error, $\tfrac{1}{N}\sum_i \lvert \hat y_i - y_i \rvert$; identical in definition to $\mathcal{L}_{\mathrm{L1}}$ of \Cref{eq:pixel_losses}, reported here as a metric on the normalized intensities.
  \item \textbf{Gradient MAE.} Edge-preservation metric: $3 \times 3$ Sobel operators are applied to $\hat Y$ and $Y$ separately in the horizontal and vertical directions, gradient magnitudes $\lVert \nabla I\rVert = \sqrt{G_x^2 + G_y^2}$ are computed for each, and the metric is the mean absolute error between the two magnitude maps.
  \item \textbf{PSNR.} $\mathrm{PSNR}(\hat Y, Y) = 10 \log_{10}\!\bigl(L^{2} / \mathrm{MSE}(\hat Y, Y)\bigr)$ in decibels, with peak intensity $L = 1$; pairs with $\mathrm{MSE} = 0$ are assigned $+\infty$ and excluded from averages.
\end{itemize}
Metrics are evaluated on every interpolated slice of every test triplet; slice-level values are averaged within each patient to form the patient-level quantities used in the statistical analysis of \Cref{sec:stats}. Because the reference middle slice $I_{k+1}$ carries acquisition noise, a model that perfectly recovers the conditional-expectation signal of \Cref{sec:noise} would differ from it by the suppressed noise; see \Cref{sec:limitations} for the implications when comparing top-performing losses.

\subsection{ROI-level noise quantification}
\label{sec:noise-methods}
We quantify the implicit-denoising property analysed in \Cref{sec:noise} via the ROI-level protocol below.
To quantify within-tissue noise in held-out test predictions, we measure the pixel standard deviation in three anatomically fixed white-matter regions of interest (ROIs) per slice: left and right centrum semiovale, and pons. Each ROI is a $16\times16$-pixel patch placed by a brain-mask-based heuristic (centroid and bounding box of the largest connected component at threshold $0.05$); ROIs that extend outside the brain mask, overlap a hemorrhage-labelled slice, or contain intensity outliers exceeding three times the patient median ROI standard deviation are rejected. Of $968 \times 3 = 2904$ candidate ROI--triplet combinations, $1262$ are retained after rejection ($396$ dropped for hemorrhage overlap, $183$ for geometry, $219$ for outliers, $18$ pons ROIs unplaceable). Within each surviving ROI we compute the pixel standard deviation $\sigma_\mathrm{ROI}$ of the acquired middle slice and of each model's prediction, and define the noise-reduction ratio $\eta = (\sigma^\mathrm{acq} - \sigma^\mathrm{pred})/\sigma^\mathrm{acq}$ (positive: the prediction has lower in-ROI variance than the acquired reference). Per-patient $\eta$ is the mean over surviving ROIs within each triplet and then over triplets within each patient; we test each model against zero with a paired Wilcoxon signed-rank across the $28$ patients with any surviving ROI, and apply Benjamini--Hochberg FDR across the 5-model family. The radially-averaged noise power spectrum (NPS) of the same ROIs, computed following ICRU~54 conventions, complements the scalar analysis by showing which spatial frequencies are suppressed.

\subsection{Statistical analysis}
\label{sec:stats}

Primary inference is \textbf{patient-level} (30 independent patients). The paired metric differences are heavy-tailed, so we use nonparametric procedures throughout:
\begin{enumerate}[leftmargin=*]
  \item compute metric means per patient;
  \item estimate mean and 95\% CI via nonparametric percentile bootstrap (10,000 resamples)~\cite{Efron1979};
  \item perform two-sided paired Wilcoxon signed-rank tests~\cite{wilcoxon1945} of each candidate against the reference model (MS-SSIM+L1, 0.5/0.5), a one-vs-reference design rather than an all-pairs omnibus, with the reference designated \emph{a priori} (rationale and candidate list in \Cref{sec:paired}).
\end{enumerate}
Patient-level pairing ensures that each model comparison tests whether one loss consistently outperforms the other across the same 30 anatomies, thereby controlling for inter-patient variance rather than relying on absolute metric magnitudes. We adopt a significance level of $\alpha = 0.05$ for all tests.

To control the false discovery rate across the patient-level paired-test family (50 tests: 10 candidate models $\times$ 5 reported metrics; full enumeration in \Cref{sec:paired}), we apply the Benjamini--Hochberg procedure~\cite{benjamini1995controlling} at $q < 0.05$; BH-adjusted $q$-values are reported alongside raw $p$-values. Slice-level analyses are also computed for transparency, but we avoid treating slice-level $p$-values as primary due to within-patient dependence.

\paragraph{Exploratory subgroup analysis.}
Separate from the primary patient-level paired comparisons, \Cref{sec:hemorrhage-strat} reports an \emph{exploratory, post-hoc} subgroup analysis testing whether slice-level metrics differ between hemorrhage and normal target slices. Because the two subgroups are different slices (not paired), we use the two-sample Mann--Whitney $U$ test (two-sided), with Benjamini--Hochberg FDR applied jointly across the $8\times3$ primary-metric hemorrhage-analysis family and a patient-level cluster bootstrap reported as an independence-aware sensitivity check; effect-size and bootstrap definitions are given in \Cref{sec:appendix_hemorrhage_tests}.
\paragraph{Resolution limits.}
No a priori power calculation was performed; reported effect sizes are observed rather than targeted. The two-sided exact Wilcoxon signed-rank test at $n = 30$ has a minimum attainable $p$-value of $1.86 \times 10^{-9}$, which bounds patient-level significance claims from below. Slice-level tests are reported for transparency but can be anti-conservative under within-patient dependence; the hemorrhage-stratification analysis therefore reports patient-level cluster-bootstrap $p$-values as an independence-aware sensitivity check.

%% file: figures/architecture.tex
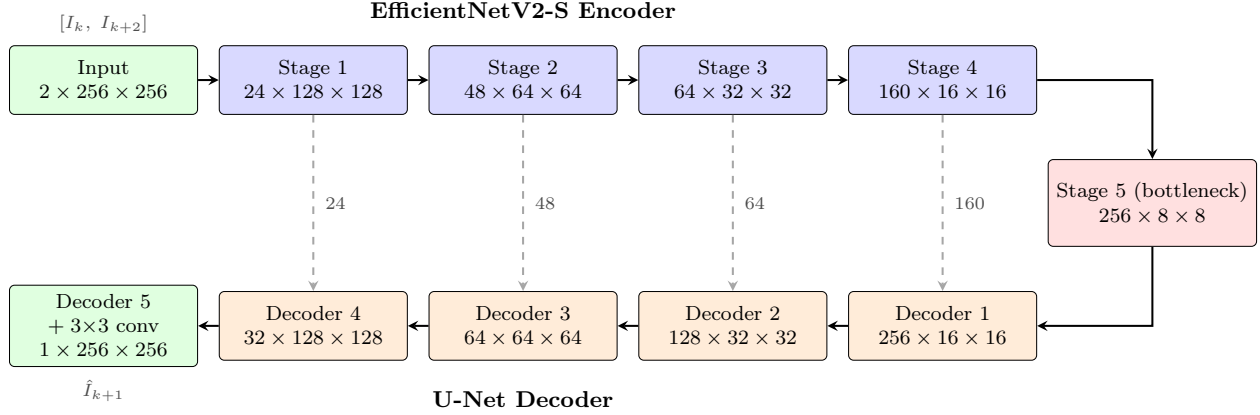
\begin{figure}[t]
  \centering
  \resizebox{\linewidth}{!}{%
  \begin{tikzpicture}[
    enc/.style={draw, fill=blue!15, minimum width=2.6cm, minimum height=0.95cm,
                font=\footnotesize, rounded corners=2pt, align=center},
    dec/.style={draw, fill=orange!15, minimum width=2.6cm, minimum height=0.95cm,
                font=\footnotesize, rounded corners=2pt, align=center},
    io/.style={draw, fill=green!12, minimum width=2.6cm, minimum height=0.95cm,
               font=\footnotesize, rounded corners=2pt, align=center},
    bottle/.style={draw, fill=red!12, minimum width=2.6cm, minimum height=1.2cm,
                   font=\footnotesize, rounded corners=2pt, align=center},
    skip/.style={->, >=stealth, dashed, thick, gray!70},
    flow/.style={->, >=stealth, thick},
    lbl/.style={font=\scriptsize, text=black!70},
  ]

    \def\cA{0}      
    \def\cB{2.9}    
    \def\cC{5.8}    
    \def\cD{8.7}    
    \def\cE{11.6}   
    \def\cF{14.5}   
    \def\rT{1.7}    
    \def\rB{-1.7}   

    \node[io]  (input) at (\cA, \rT) {Input\\$2 \times 256 \times 256$};
    \node[enc] (e1)    at (\cB, \rT) {Stage 1\\$24 \times 128 \times 128$};
    \node[enc] (e2)    at (\cC, \rT) {Stage 2\\$48 \times 64 \times 64$};
    \node[enc] (e3)    at (\cD, \rT) {Stage 3\\$64 \times 32 \times 32$};
    \node[enc] (e4)    at (\cE, \rT) {Stage 4\\$160 \times 16 \times 16$};

    \node[bottle] (bn) at (\cF, 0) {Stage 5 (bottleneck)\\$256 \times 8 \times 8$};

    \node[io]  (output) at (\cA, \rB) {Decoder 5\\$+\;3{\times}3$ conv\\$1 \times 256 \times 256$};
    \node[dec] (d4)     at (\cB, \rB) {Decoder 4\\$32 \times 128 \times 128$};
    \node[dec] (d3)     at (\cC, \rB) {Decoder 3\\$64 \times 64 \times 64$};
    \node[dec] (d2)     at (\cD, \rB) {Decoder 2\\$128 \times 32 \times 32$};
    \node[dec] (d1)     at (\cE, \rB) {Decoder 1\\$256 \times 16 \times 16$};

    \draw[flow] (input) -- (e1);
    \draw[flow] (e1) -- (e2);
    \draw[flow] (e2) -- (e3);
    \draw[flow] (e3) -- (e4);
    \draw[flow] (e4.east) -| (bn.north);
    \draw[flow] (bn.south) |- (d1.east);
    \draw[flow] (d1) -- (d2);
    \draw[flow] (d2) -- (d3);
    \draw[flow] (d3) -- (d4);
    \draw[flow] (d4) -- (output);

    \draw[skip] (e4.south) -- node[lbl, right=1pt] {160} (d1.north);
    \draw[skip] (e3.south) -- node[lbl, right=1pt] {64}  (d2.north);
    \draw[skip] (e2.south) -- node[lbl, right=1pt] {48}  (d3.north);
    \draw[skip] (e1.south) -- node[lbl, right=1pt] {24}  (d4.north);

    \node[font=\small\bfseries, above=0.25cm] at ($(input.north)!0.5!(e4.north)$)
      {EfficientNetV2-S Encoder};
    \node[font=\small\bfseries, below=0.25cm] at ($(output.south)!0.5!(d1.south)$)
      {U-Net Decoder};

    \node[lbl, above=0.05cm] at (input.north) {$[I_k,\; I_{k+2}]$};
    \node[lbl, below=0.05cm] at (output.south) {$\hat{I}_{k+1}$};

  \end{tikzpicture}%
  }
  \caption{U-Net + EfficientNetV2-S architecture for CT slice interpolation.
    Top row: encoder (blue), left-to-right in decreasing spatial resolution.
    Bottom row: decoder (orange), right-to-left in increasing spatial
    resolution. Columns are aligned by spatial resolution: the leftmost
    column carries the 2-channel input $[I_k, I_{k+2}]$ and the final
    decoder block fused with its $3 \times 3$ conv output projection
    (producing the single-channel prediction $\hat{I}_{k+1}$); the next four
    columns pair each encoder stage with the decoder block that consumes its
    skip connection (dashed, annotated with the skip channel count); the
    rightmost column carries the shared bottleneck. The absence of a skip
    arrow in the leftmost column reflects the fact that the final decoder
    block has no encoder skip. All dimensions shown correspond to the
    $256 \times 256$ patch resolution used at both training and inference
    time; at test time, 13 such patch predictions are reassembled into a
    $512 \times 512$ composite (see \Cref{sec:model}).}
  \label{fig:architecture}
\end{figure}

%% file: figures/augmentation.tex
%
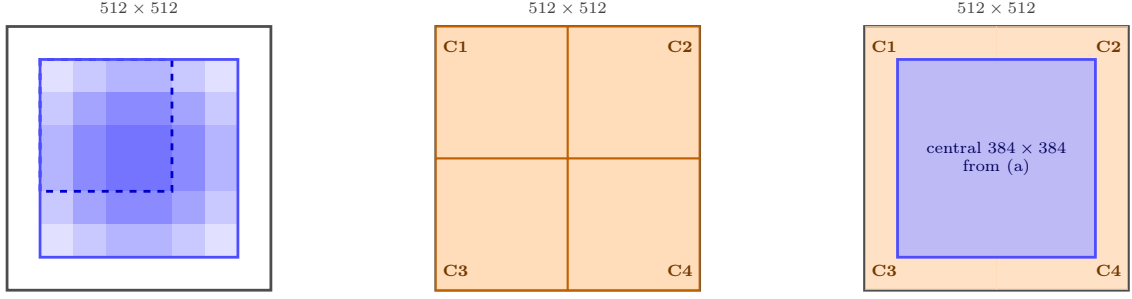
\begin{figure}[t]
  \centering
  \resizebox{\linewidth}{!}{%
  \begin{tikzpicture}[
    slice/.style={draw, very thick, black!70},
    centerband/.style={draw=blue!70, very thick},
    cropfill/.style={fill=blue!65, fill opacity=0.18},
    highlightcrop/.style={draw=blue!80!black, very thick, dashed},
    corner/.style={draw=orange!75!black, thick, fill=orange!30, fill opacity=0.55},
    lblbig/.style={font=\small\bfseries},
    lbl/.style={font=\scriptsize, text=black!75},
    tag/.style={font=\scriptsize\bfseries, text=orange!45!black},
  ]

    \begin{scope}[xshift=0cm]
      \draw[slice] (0,0) rectangle (4,4);
      \node[lbl, above=0.05cm] at (2, 4) {$512 \times 512$};

      \foreach \bx/\by in {0.5/1.5, 1.0/1.5, 1.5/1.5,
                           0.5/1.0, 1.0/1.0, 1.5/1.0,
                           0.5/0.5, 1.0/0.5, 1.5/0.5} {
        \path[cropfill] (\bx, \by) rectangle (\bx + 2, \by + 2);
      }

      \draw[highlightcrop] (0.5, 1.5) rectangle (2.5, 3.5);

      \draw[centerband] (0.5, 0.5) rectangle (3.5, 3.5);

      \node[lblbig, below=1.2cm] at (2, 0)
        {(a) 9 overlapping center crops};
    \end{scope}

    \begin{scope}[xshift=6.5cm]
      \draw[slice] (0,0) rectangle (4,4);
      \node[lbl, above=0.05cm] at (2, 4) {$512 \times 512$};

      \fill[orange!30, fill opacity=0.75] (0, 0) rectangle (2, 2);
      \fill[orange!30, fill opacity=0.75] (2, 0) rectangle (4, 2);
      \fill[orange!30, fill opacity=0.75] (0, 2) rectangle (2, 4);
      \fill[orange!30, fill opacity=0.75] (2, 2) rectangle (4, 4);
      \draw[corner] (0, 0) rectangle (2, 2);
      \draw[corner] (2, 0) rectangle (4, 2);
      \draw[corner] (0, 2) rectangle (2, 4);
      \draw[corner] (2, 2) rectangle (4, 4);

      \node[tag] at (0.3, 3.7)  {C1};
      \node[tag] at (3.7, 3.7)  {C2};
      \node[tag] at (0.3, 0.3)  {C3};
      \node[tag] at (3.7, 0.3)  {C4};

      \node[lblbig, below=1.2cm] at (2, 0)
        {(b) 4 corner patches};
    \end{scope}

    \begin{scope}[xshift=13cm]
      \draw[slice] (0,0) rectangle (4,4);
      \node[lbl, above=0.05cm] at (2, 4) {$512 \times 512$};

      \fill[orange!30, fill opacity=0.75] (0, 0) rectangle (2, 2);
      \fill[orange!30, fill opacity=0.75] (2, 0) rectangle (4, 2);
      \fill[orange!30, fill opacity=0.75] (0, 2) rectangle (2, 4);
      \fill[orange!30, fill opacity=0.75] (2, 2) rectangle (4, 4);

      \fill[blue!30, fill opacity=0.85] (0.5, 0.5) rectangle (3.5, 3.5);
      \draw[centerband] (0.5, 0.5) rectangle (3.5, 3.5);

      \node[align=center, font=\scriptsize, text=blue!35!black] at (2, 2)
        {central $384 \times 384$\\from (a)};
      \node[tag] at (0.3, 3.7)  {C1};
      \node[tag] at (3.7, 3.7)  {C2};
      \node[tag] at (0.3, 0.3)  {C3};
      \node[tag] at (3.7, 0.3)  {C4};

      \node[lblbig, below=1.2cm] at (2, 0)
        {(c) Composite: (a) over (b), 13 patches};
    \end{scope}

  \end{tikzpicture}%
  }
  \caption{Patch geometry shared by training-time sampling and test-time
    reconstruction. (a) Nine overlapping $256 \times 256$ crops whose
    top-left corners form a $3\times 3$ grid at 64-pixel stride in the
    $512\times 512$ slice; their union covers the central $384\times 384$
    band (solid blue outline). Translucent fills overlap so deeper color
    indicates regions covered by more crops; the dashed outline marks one
    representative crop (top-left of the grid) to indicate the
    $256\times 256$ scale of a single view. At training time one crop is
    drawn per sample; at test time their predictions are averaged over
    overlaps to reconstruct the central band. (b) Four non-overlapping $256\times 256$
    corner patches $C_1$--$C_4$ with top-left corners at $(y,x) = (0,0)$,
    $(0,256)$, $(256,0)$, $(256,256)$, tiling the full $512\times 512$ slice. (c) Test-time
    composite: the central $384\times 384$ band from (a) is placed on top
    of the four corner patches from (b), yielding a 13-patch $512\times 512$
    prediction that is directly comparable to the classical and VFI
    baselines.}
  \label{fig:augmentation}
\end{figure}

%% file: sections/results.tex
\section{Results}
\label{sec:results}

\subsection{Held-out test performance (patient-level primary analysis)}
\label{sec:test-results}
All four primary loss families (SSIM$_{\mathrm{s}}$, L1, MS-SSIM+L1, MSE) converged within the 500-epoch budget; the full configuration registry, including per-run learning rates and termination epochs, is in Appendix~\ref{sec:appendix_registry}. Validation metrics on $256 \times 256$ crops are not directly comparable to the test metrics on the reassembled $512 \times 512$ composite, whose outer $64$-pixel near-zero border inflates structural scores; we therefore report the held-out test analysis as the primary comparison.

The primary analysis is conducted on the 30-patient held-out test set: per-patient metric means are computed within each patient and then summarized across patients with 95\% bootstrap confidence intervals under the protocol of \Cref{sec:stats}. \Cref{tab:test_patient} reports five per-image metrics (SSIM, MS-SSIM, MAE, gradient MAE, and PSNR) for the two classical baselines, the two pretrained VFI models, and four learned configurations; the paragraphs that follow group these rows into baselines, VFI transfer, and the controlled versus separately tuned learned runs.

\begin{table}[t]
  \centering
  \caption{Primary test-set results (patient-level, 30 patients). Three learned models (L1, MS-SSIM+L1, MSE) share identical optimization settings (lr $= 8\!\times\!10^{-4}$, batch size 96) and constitute the controlled comparison. The subscript~\textit{s} on SSIM$_{\mathrm{s}}$ denotes the training-stability variant ($K_2 = 0.4$) run with a separate hyperparameter setting; see \Cref{sec:ssim-fragility}. Bold marks the per-column best; ties at the third decimal place are not bolded. The reference is the noisy acquired middle slice; small inter-model deltas may reflect noise-matching rather than signal recovery (see \Cref{sec:limitations}).}
  \label{tab:test_patient}
  \footnotesize
  \setlength{\tabcolsep}{4pt}
  \resizebox{\linewidth}{!}{%
  \begin{tabular}{lccccc}
    \toprule
    Method & SSIM $\uparrow$ & MS-SSIM $\uparrow$ & MAE $\downarrow$ & Grad-MAE $\downarrow$ & PSNR $\uparrow$ \\
    \midrule
    Cubic interpolation baseline & 0.778 [0.766, 0.791] & 0.823 [0.815, 0.831] & 0.0432 [0.0417, 0.0448] & 0.116 [0.110, 0.122] & 19.051 [18.766, 19.332] \\
    Mean interpolation baseline & 0.784 [0.773, 0.796] & 0.822 [0.815, 0.830] & 0.0436 [0.0421, 0.0452] & 0.112 [0.107, 0.118] & 19.025 [18.753, 19.284] \\
    \midrule
    RIFE~\cite{RIFE_ECCV2022} & 0.847 [0.839, 0.855] & 0.887 [0.882, 0.893] & 0.0262 [0.0249, 0.0276] & 0.0798 [0.0757, 0.0841] & 22.307 [21.849, 22.773] \\
    FILM~\cite{reda2022film} & 0.848 [0.840, 0.856] & 0.891 [0.885, 0.896] & 0.0255 [0.0242, 0.0269] & 0.0817 [0.0774, 0.0860] & 22.425 [21.935, 22.912] \\
    \midrule
    \multicolumn{6}{l}{\textit{Controlled comparison (lr $= 8\!\times\!10^{-4}$, batch size 96)}} \\
    L1 & 0.890 [0.881, 0.899] & 0.936 [0.931, 0.941] & \textbf{0.0166} [0.0154, 0.0178] & 0.0717 [0.0666, 0.0766] & 25.402 [24.773, 26.057] \\
    MS-SSIM+L1 (ref.) & 0.888 [0.879, 0.897] & 0.936 [0.931, 0.941] & 0.0172 [0.0160, 0.0184] & 0.0726 [0.0678, 0.0774] & \textbf{25.435} [24.864, 26.030] \\
    MSE & 0.871 [0.862, 0.881] & 0.931 [0.926, 0.937] & 0.0199 [0.0186, 0.0212] & 0.0774 [0.0724, 0.0824] & 25.399 [24.854, 25.972] \\
    \cmidrule(lr){1-6}
    \multicolumn{6}{l}{\textit{Separately tuned; controlled SSIM run at lr $= 8\!\times\!10^{-4}$ catastrophically failed (test SSIM $= 0.067$; see \Cref{sec:ssim-fragility})}} \\
    SSIM$_{\mathrm{s}}$ (lr $= 3\!\times\!10^{-3}$, bs 32) & 0.890 [0.881, 0.899] & 0.935 [0.930, 0.940] & 0.0192 [0.0180, 0.0204] & \textbf{0.0714} [0.0669, 0.0760] & 24.955 [24.401, 25.510] \\
    \bottomrule
  \end{tabular}%
  }
\end{table}

Three baseline categories are included: two classical methods (pixel-wise mean and cubic spline interpolation along the z-axis), and two pretrained VFI models applied directly to CT without fine-tuning (RIFE~\cite{RIFE_ECCV2022} and FILM~\cite{reda2022film}). The classical baselines represent naive interpolation and establish a floor rather than a clinical standard of care. RIFE and FILM, trained on natural video, sit between the classical and learned models on structural metrics (SSIM $0.847$--$0.848$, vs.\ ${\sim}0.78$ classical and ${\sim}0.89$ learned), confirming that natural-video priors transfer partially but not fully to CT. Numerically, every domain-specific model in \Cref{tab:test_patient} exceeds both baseline categories on SSIM and MS-SSIM; in the patient-level paired analysis, the reference model improves over all four baselines on both structural metrics at the Wilcoxon floor ($p = 1.86\!\times\!10^{-9}$).

Among the learned models, three (L1, MS-SSIM+L1, MSE) share identical optimization settings (lr $= 8\!\times\!10^{-4}$, batch size 96) and constitute the controlled comparison. The SSIM$_{\mathrm{s}}$ model used a higher learning rate ($3\!\times\!10^{-3}$) and smaller batch size (32); its results are included because they demonstrate what SSIM loss can achieve under favorable hyperparameters, but they represent a separately tuned configuration rather than a controlled data point (see \Cref{sec:ssim-fragility}).

No single loss dominates all metrics within the controlled subset: L1 leads MAE, MS-SSIM+L1 has the numerically highest PSNR, and L1 and MS-SSIM+L1 tie on MS-SSIM. The separately tuned SSIM$_{\mathrm{s}}$ achieves the lowest (best) gradient MAE, with L1 effectively tied for the highest SSIM at the third decimal.

\subsection{Paired tests vs reference model}
\label{sec:paired}
Following the patient-level protocol of \Cref{sec:stats}, we test whether any candidate model (classical, VFI, or alternative learned-loss configuration) produces a patient-level improvement over the reference model, $\mathcal{L}_{\mathrm{MS\text{-}SSIM+L1}}$ at $\alpha = 0.5$. This reference was designated \emph{a priori} with respect to the test set: the MS-SSIM$+$L1 family itself follows the hybrid perceptual$+$pixel recipe of Zhao \emph{et al.}~\cite{Zhao2017LossFF}, and among configurations trained under the controlled hyperparameter setting (lr $=8\!\times\!10^{-4}$, batch size 96), the $\alpha = 0.5$ weighting trained stably to early stopping and was retained as the controlled-comparison candidate (full registry in Appendix~\ref{sec:appendix_registry}). No test-set metric was used in selecting the reference. We report patient-level paired improvements (positive values mean better candidate under each metric direction convention).

The 10 candidates comprise 4 baselines (pixel-wise mean, cubic z-interpolation, RIFE, FILM) and 6 non-reference learned runs: SSIM$_{\mathrm{s}}$, L1, MSE, the catastrophically failed controlled SSIM run at lr $= 8\!\times\!10^{-4}$, a lower-learning-rate L1 configuration, and a second MS-SSIM+L1 weighting. The reference model itself is excluded from the candidate list. The five reported metrics of \Cref{sec:metrics} (SSIM, MS-SSIM, MAE, gradient MAE, PSNR) are evaluated, yielding $10 \times 5 = 50$ tests. \Cref{tab:paired_patient} reports a representative subset (SSIM, MAE, PSNR); the full patient-level 50-test table with BH-adjusted $q$-values is released as a tabular artefact under \Cref{sec:appendix_repro}.

\begin{table}[t]
  \centering
  \caption{Patient-level paired improvements vs.\ reference (MS-SSIM+L1). Positive $\Delta$ = candidate better. Raw two-sided paired Wilcoxon signed-rank $p$-values are shown; Benjamini--Hochberg FDR correction over the full 50-test family leaves every significance decision at $p<0.05$ unchanged (see \Cref{sec:stats}). Bootstrap CI endpoints use 10{,}000 resamples. The full patient-level 50-test table is released as a tabular artefact under \Cref{sec:appendix_repro}, and contains, e.g., the paired MS-SSIM test for L1 vs.\ the reference (narrowly significant at $p = 0.0087$), not tabulated here.}
  \label{tab:paired_patient}
  \footnotesize
  \resizebox{\linewidth}{!}{%
  \begin{tabular}{lcccccc}
    \toprule
    Candidate & $\Delta$SSIM & $p$ & $\Delta$MAE & $p$ & $\Delta$PSNR & $p$ \\
    \midrule
    Cubic baseline & $-$0.1096 [$-$0.1145, $-$0.1048] & $1.86\!\times\!10^{-9}$ & $-$0.0260 [$-$0.0272, $-$0.0248] & $1.86\!\times\!10^{-9}$ & $-$6.384 [$-$6.846, $-$5.935] & $1.86\!\times\!10^{-9}$ \\
    Mean baseline & $-$0.1039 [$-$0.1082, $-$0.0998] & $1.86\!\times\!10^{-9}$ & $-$0.0264 [$-$0.0276, $-$0.0252] & $1.86\!\times\!10^{-9}$ & $-$6.410 [$-$6.892, $-$5.931] & $1.86\!\times\!10^{-9}$ \\
    \midrule
    RIFE~\cite{RIFE_ECCV2022} & $-$0.0405 [$-$0.0436, $-$0.0377] & $1.86\!\times\!10^{-9}$ & $-$0.0090 [$-$0.0097, $-$0.0084] & $1.86\!\times\!10^{-9}$ & $-$3.128 [$-$3.452, $-$2.831] & $1.86\!\times\!10^{-9}$ \\
    FILM~\cite{reda2022film} & $-$0.0396 [$-$0.0424, $-$0.0370] & $1.86\!\times\!10^{-9}$ & $-$0.0083 [$-$0.0090, $-$0.0077] & $1.86\!\times\!10^{-9}$ & $-$3.010 [$-$3.278, $-$2.747] & $1.86\!\times\!10^{-9}$ \\
    \midrule
    SSIM$_{\mathrm{s}}$ & $+$0.0017 [0.0011, 0.0023] & $1.19\!\times\!10^{-6}$ & $-$0.0020 [$-$0.0021, $-$0.0019] & $1.86\!\times\!10^{-9}$ & $-$0.479 [$-$0.561, $-$0.406] & $1.86\!\times\!10^{-9}$ \\
    L1 & $+$0.0021 [0.0018, 0.0025] & $1.86\!\times\!10^{-9}$ & $+$0.0006 [0.0005, 0.0007] & $1.86\!\times\!10^{-9}$ & $-$0.033 [$-$0.119, 0.069] & 0.171 \\
    MSE & $-$0.0170 [$-$0.0184, $-$0.0157] & $1.86\!\times\!10^{-9}$ & $-$0.0027 [$-$0.0029, $-$0.0025] & $1.86\!\times\!10^{-9}$ & $-$0.035 [$-$0.141, 0.051] & 0.761 \\
    \bottomrule
  \end{tabular}%
  }
\end{table}

The majority of p-values fall at the Wilcoxon floor for $n=30$ pairs ($1.86 \times 10^{-9}$), so Benjamini--Hochberg FDR correction (see \Cref{sec:stats}) leaves every significance decision at $p<0.05$ unchanged. The few non-significant comparisons are small learned-model differences, including PSNR for models with near-identical performance (e.g., MSE PSNR $p = 0.761$, L1 PSNR $p = 0.171$) and MS-SSIM contrasts omitted from \Cref{tab:paired_patient}. Because paired testing can detect very small but consistent directional differences, we report confidence intervals alongside p-values so that readers can judge practical significance from the effect magnitudes rather than from p-values alone.

\subsection{Hemorrhage stratification}
\label{sec:hemorrhage-strat}
Recall that the 30-patient test set is enriched for hemorrhage subtypes (\Cref{sec:data}); the slice-level prevalence below (41\% hemorrhage) does not reflect unselected clinical cohorts. \Cref{tab:hemorrhage_strat} stratifies test-set performance by hemorrhage presence at the slice level, using the RSNA \texttt{any}-subtype label on the target middle slice $I_{k+1}$ of each triplet. Hemorrhage slices ($n=396$) yield lower SSIM than normal slices ($n=572$) across every evaluated method (classical, VFI, and learned) with a typical gap of $0.02$--$0.03$.

\begin{table}[t]
  \centering
  \caption{Slice-level performance stratified by hemorrhage presence (396 hemorrhage, 572 normal triplets). $\uparrow$~higher is better; $\downarrow$~lower is better. Best value per column in \textbf{bold}. Bootstrap 95\% CIs are slice-level and understate true patient-level uncertainty; they are provided for descriptive completeness. Inferential tests are reported separately in \Cref{tab:hemorrhage_tests} (\Cref{sec:appendix_hemorrhage_tests}).}
  \label{tab:hemorrhage_strat}
  \footnotesize
  \begin{tabular}{lcccc}
    \toprule
    & \multicolumn{2}{c}{SSIM $\uparrow$} & \multicolumn{2}{c}{MAE $\downarrow$} \\
    \cmidrule(lr){2-3} \cmidrule(lr){4-5}
    Method & Hemorrhage & Normal & Hemorrhage & Normal \\
    \midrule
    Cubic baseline & 0.759 [0.753, 0.765] & 0.789 [0.782, 0.796] & 0.039 [0.038, 0.041] & 0.046 [0.045, 0.047] \\
    Mean baseline & 0.768 [0.762, 0.773] & 0.792 [0.786, 0.799] & 0.039 [0.038, 0.040] & 0.047 [0.046, 0.048] \\
    \midrule
    RIFE baseline & 0.827 [0.823, 0.832] & 0.859 [0.853, 0.865] & 0.024 [0.023, 0.026] & 0.028 [0.026, 0.029] \\
    FILM baseline & 0.828 [0.823, 0.833] & 0.861 [0.854, 0.867] & 0.024 [0.023, 0.025] & 0.027 [0.025, 0.028] \\
    \midrule
    SSIM$_{\mathrm{s}}$ & \textbf{0.873} [0.869, 0.877] & 0.899 [0.894, 0.904] & 0.020 [0.019, 0.020] & 0.019 [0.018, 0.020] \\
    L1 & \textbf{0.873} [0.868, 0.877] & \textbf{0.900} [0.895, 0.904] & \textbf{0.017} [0.016, 0.018] & \textbf{0.017} [0.016, 0.018] \\
    MS-SSIM+L1 (ref.) & 0.871 [0.866, 0.875] & 0.897 [0.892, 0.902] & 0.018 [0.017, 0.018] & \textbf{0.017} [0.016, 0.018] \\
    MSE & 0.856 [0.851, 0.860] & 0.879 [0.873, 0.884] & 0.020 [0.019, 0.021] & 0.020 [0.019, 0.021] \\
    \bottomrule
  \end{tabular}
\end{table}

We tested this gap inferentially using a two-sided Mann--Whitney $U$ test per (method, metric) cell over the 8 methods and 3 primary metrics (SSIM, MAE, PSNR), a patient-level cluster bootstrap ($p_\mathrm{cb}$; 10{,}000 resamples) to account for within-patient dependence, and Benjamini--Hochberg FDR control~\cite{benjamini1995controlling} over the full $8\times 3 = 24$-test family. Full methodology, effect-size definitions, and $q$-values are in \Cref{sec:appendix_hemorrhage_tests} (\Cref{tab:hemorrhage_tests}). The analysis was designed after seeing \Cref{tab:hemorrhage_strat} and is therefore post-hoc / exploratory.

The SSIM gap (hemorrhage $<$ normal) is unanimous across the 8 methods and survives the cluster bootstrap in 7 of 8 cases at $p_\mathrm{cb}\le 0.018$; the Mean baseline is marginal ($p_\mathrm{cb}=0.056$). For MAE and PSNR the test family splits by method class: classical baselines show hemorrhage-normal gaps significant under both the asymptotic and cluster-bootstrap $p$; VFI baselines show the same pattern for PSNR but not MAE; and the four learned models show small effects ($|r|\le 0.10$) that are marginally significant asymptotically but do \emph{not} survive the patient-level bootstrap ($p_\mathrm{cb}\in[0.14, 0.34]$). We therefore interpret the hemorrhage-vs-normal effect in learned models as an SSIM phenomenon; the mechanism is discussed in \Cref{sec:loss-pathology}.

\subsection{Implicit denoising: noise reduction in uniform ROIs}
\label{sec:noise-results}
\begin{figure}[t]
  \centering
  \includegraphics[width=0.85\linewidth,keepaspectratio]{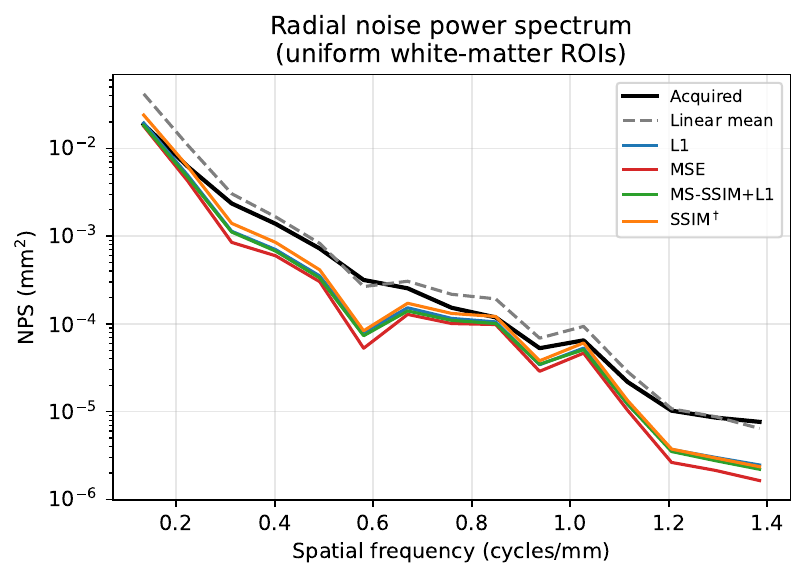}
  \caption{Radially-averaged 1D noise power spectrum in uniform white-matter ROIs, averaged over $28$ test patients (median pixel spacing $0.488$\,mm/px). Curves compare the acquired-slice NPS to the prediction NPS of each model on the same ROIs. Pixel-wise regression losses (L1, MSE, MS-SSIM+L1) suppress NPS across essentially the full frequency band; SSIM$_{\mathrm{s}}$ shows a smaller but qualitatively similar reduction. The \texttt{linear\_mean} negative control adds variance.}
  \label{fig:nps_curves}
\end{figure}

The conditional-expectation reading of regression losses (\Cref{sec:noise}) predicts that the pixel-wise losses should suppress acquisition noise that is unpredictable from the input slices. Following the protocol of \Cref{sec:noise-methods}, we measured within-tissue noise in three anatomically fixed white-matter ROIs (left and right centrum semiovale, pons) on the held-out test set. \Cref{tab:roi_noise} summarises the patient-level results. The three pixel-wise regression losses (L1, MSE, and MS-SSIM+L1) achieve median $\eta$ of $26.0$--$26.3\%$ with tight $95\%$ patient-bootstrap CIs well above zero, and all survive BH correction at $q \le 2.9\!\times\!10^{-6}$. SSIM$_{\mathrm{s}}$ denoises less aggressively (median $\eta = 18.8\%$, CI $[11.1, 21.1]\%$, $q = 1.3\!\times\!10^{-3}$) but still significantly, consistent with SSIM's local-variance terms preserving part of the acquisition-noise texture by construction. The negative control \texttt{linear\_mean} is the arithmetic midpoint of the two neighbour slices: although simple averaging can reduce independent noise when anatomy is aligned, it is not anatomy-aware in the through-plane direction. Here it shows median $\eta = -5.3\%$, adding in-ROI variance relative to the acquired reference, as expected from averaging across inter-slice anatomical variation. The BH-significant result here confirms the test's direction sensitivity rather than denoising. \Cref{fig:nps_curves} shows that the pixel-wise regression models suppress NPS across essentially the entire frequency band in the uniform ROIs; SSIM$_{\mathrm{s}}$ shows a smaller, qualitatively similar suppression profile.

\begin{table}[t]
  \centering
  \caption{Quantitative noise reduction in uniform white-matter ROIs ($n = 28$ patients with surviving ROIs; $1262$ ROI observations). $\eta = (\sigma^\mathrm{acq} - \sigma^\mathrm{pred})/\sigma^\mathrm{acq}$: positive indicates lower in-ROI variance than the acquired reference. $95\%$ CI from patient-level bootstrap ($10{,}000$ resamples). Raw $p$ from two-sided paired Wilcoxon signed-rank; $q$ from Benjamini--Hochberg FDR across the 5-model family. \texttt{linear\_mean} is the midpoint-average negative control and is observed to add variance in this analysis.}
  \label{tab:roi_noise}
  \small
  \begin{tabular}{lrrrr}
  \toprule
  Model & Median $\eta$ (\%) & $95\%$ CI (\%) & Raw $p$ & BH $q$ \\
  \midrule
  L1                              & $+26.3$ & $[+15.2, +38.6]$ & $7.45 \times 10^{-8}$ & $3.73 \times 10^{-7}$ \\
  MSE                             & $+26.0$ & $[+13.5, +38.9]$ & $2.29 \times 10^{-6}$ & $2.86 \times 10^{-6}$ \\
  MS-SSIM+L1 (ref.)               & $+26.1$ & $[+15.7, +35.4]$ & $1.54 \times 10^{-6}$ & $2.86 \times 10^{-6}$ \\
  SSIM$_{\mathrm{s}}$ & $+18.8$ & $[+11.1, +21.1]$ & $1.29 \times 10^{-3}$ & $1.29 \times 10^{-3}$ \\
  \texttt{linear\_mean} (neg.\ control) & $-5.3$ & $[-22.6, +4.1]$ & $1.88 \times 10^{-6}$ & $2.86 \times 10^{-6}$ \\
  \bottomrule
  \end{tabular}
\end{table}

\subsection{Qualitative examples}
\label{sec:qualitative}
\begin{figure}[t]
  \centering
  \includegraphics[width=0.95\linewidth]{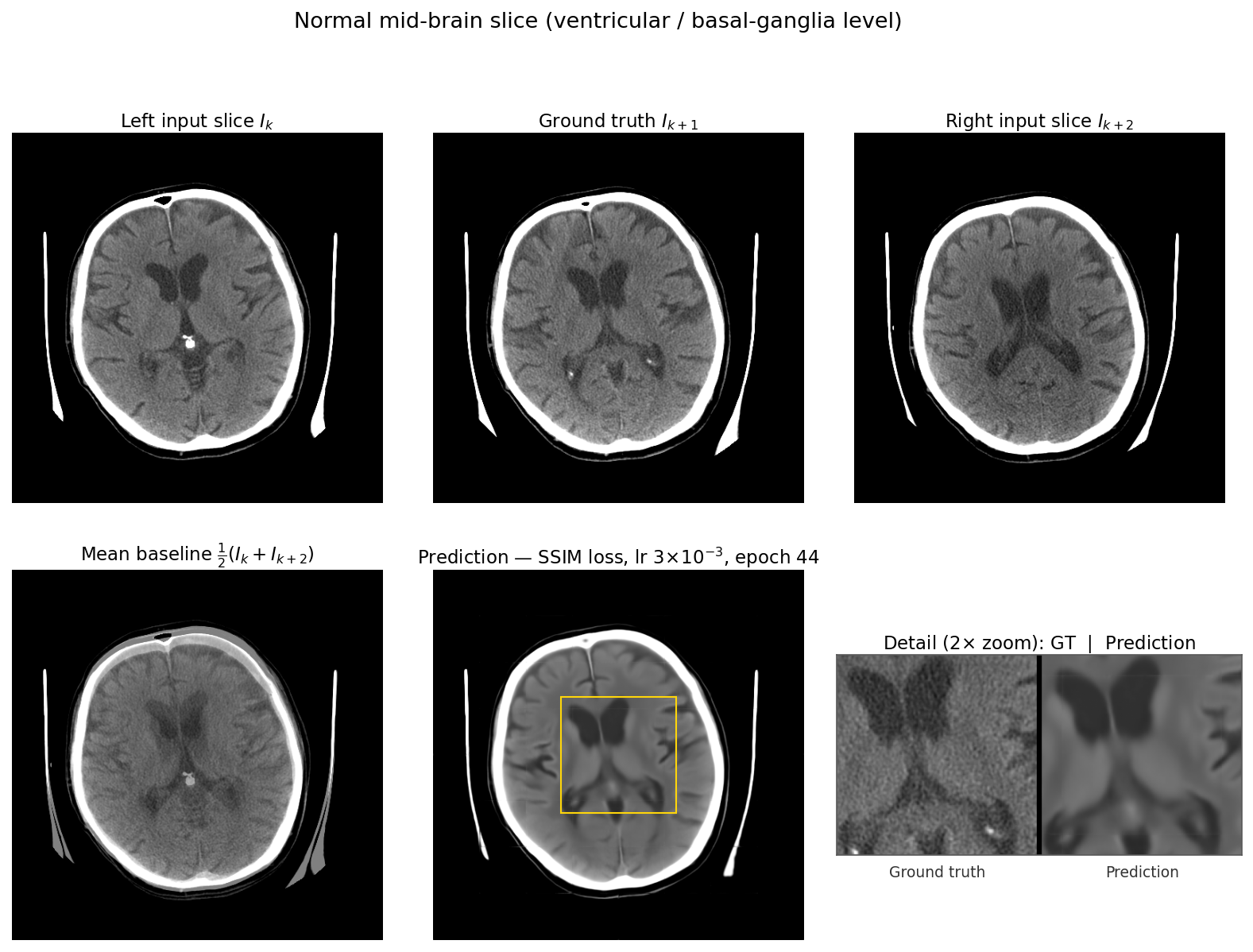}
  \caption{Qualitative interpolation result at the mid-brain ventricular / basal-ganglia level. Patient \textsf{ID\_615f69e3}, target triplet index 14; RSNA label \emph{any$=0$}. Model: SSIM loss (lr $= 3\!\times\!10^{-3}$). Detail cell: $2\times$ zoom of a $160\times160$ basal-ganglia / ventricular ROI; the yellow rectangle marks the zoom origin. Brain window and radiological convention as described in the text. The posterior-right low-intensity halo is the head-support artefact discussed in \Cref{sec:limitations}. The predicted slice is visibly smoother than the acquired reference; this is the implicit-denoising property analysed in \Cref{sec:noise} and quantified at the ROI level in \Cref{sec:noise-results}.}
  \label{fig:qual_main}
\end{figure}

\begin{figure}[p]
  \centering
  \begin{subfigure}[b]{0.71\linewidth}
    \includegraphics[width=\linewidth]{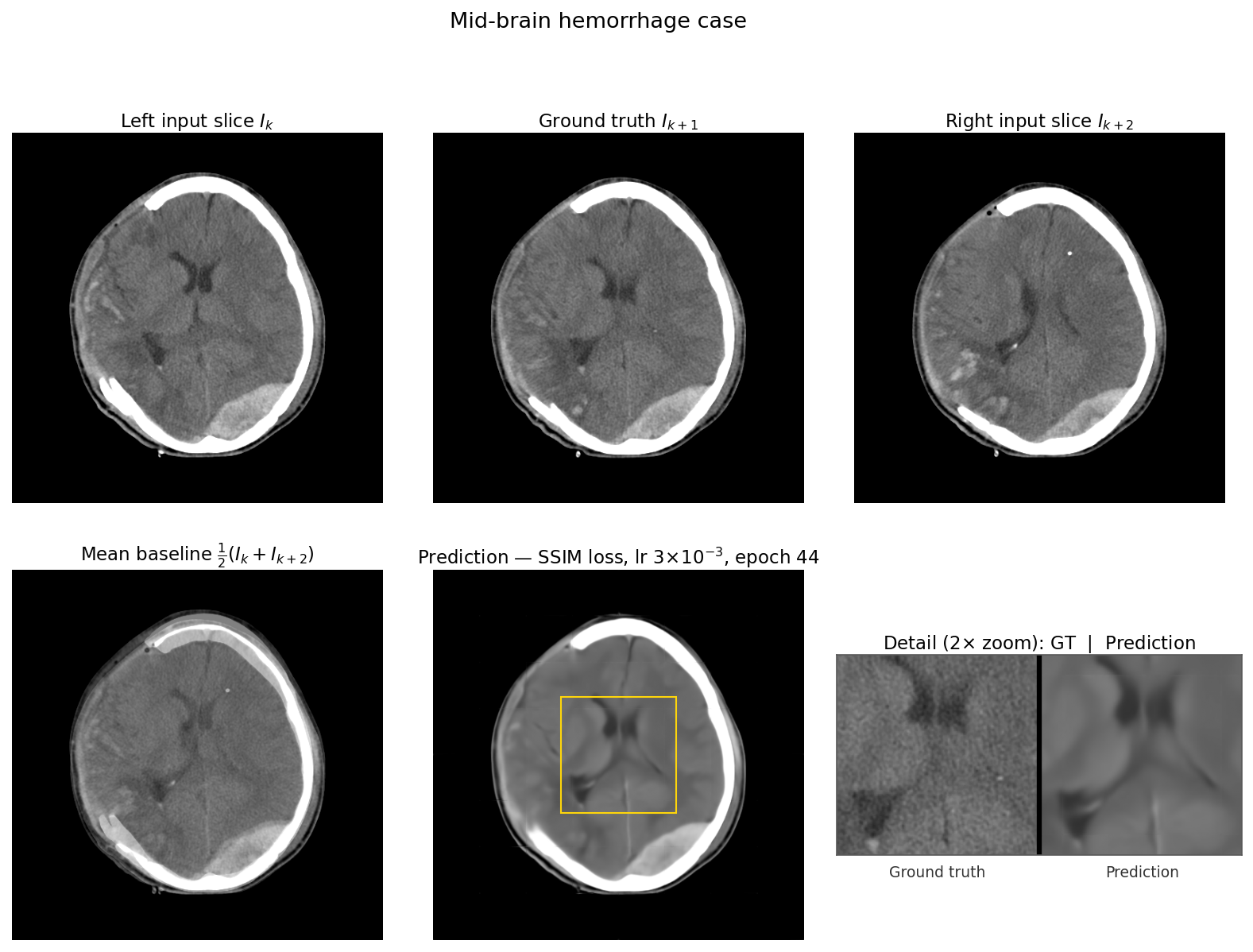}
    \caption{Mid-brain hemorrhage case; target slice carries positive RSNA labels for multiple hemorrhage subtypes (including intraparenchymal, visible as the right-sided hyperintense focus).}
  \end{subfigure}

  \vspace{0.5em}

  \begin{subfigure}[b]{0.71\linewidth}
    \includegraphics[width=\linewidth]{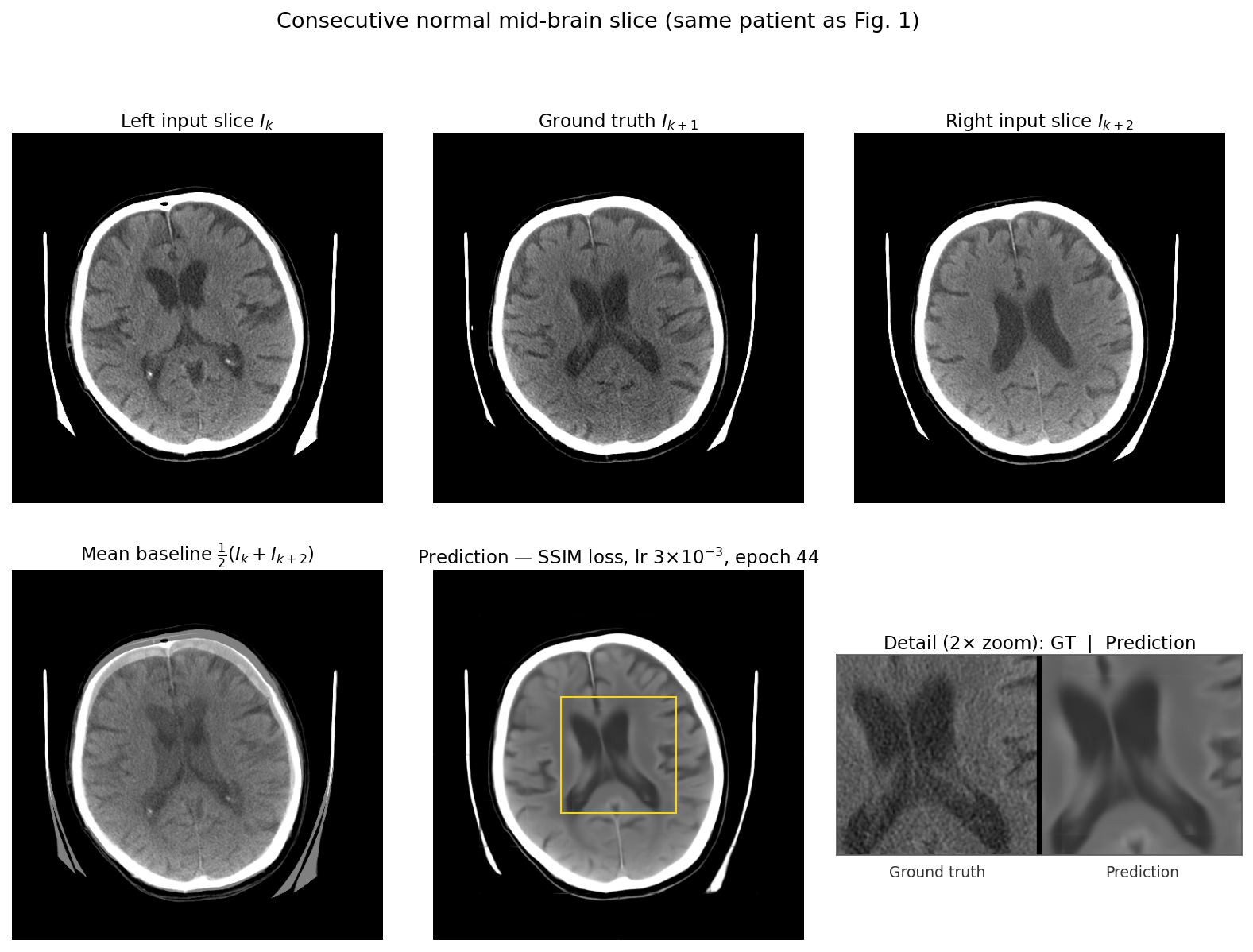}
    \caption{Consecutive mid-brain normal slice from the same patient as \Cref{fig:qual_main} (\textsf{ID\_615f69e3}; RSNA \emph{any$=0$}).}
  \end{subfigure}
  \caption{Additional qualitative examples at mid-brain axial levels: hemorrhage case (top) and consecutive normal slice (bottom). Model, display window, zoom layout, and orientation follow \Cref{fig:qual_main}. The posterior-right halo is the head-support artefact discussed in \Cref{sec:limitations}. The predicted slice is visibly smoother than the acquired reference; this is the implicit-denoising property analysed in \Cref{sec:noise} and quantified at the ROI level in \Cref{sec:noise-results}.}
  \label{fig:qual_hemorrhage}
\end{figure}

\Cref{fig:qual_main,fig:qual_hemorrhage} show representative outputs from the model that achieved the highest test-set SSIM: the SSIM loss at learning rate $3\times10^{-3}$ (SSIM$_{\mathrm{s}}$, epoch 44). Each panel shows, with in-image titles identifying every cell: top row, left input slice $I_{k}$, ground-truth middle slice $I_{k+1}$, right input slice $I_{k+2}$; bottom row, pixel-wise mean of the two input slices (mean baseline), predicted interpolation $\hat{I}_{k+1}$, and a $2\times$ zoom of a $160\times160$ region of interest shown for the ground truth and the prediction side by side. The zoom region is centred on the basal-ganglia / ventricular band and is indicated by a yellow locator rectangle drawn on the prediction panel, so that small-scale structural differences between ground truth and prediction can be assessed directly rather than inferred from the $512\times512$ views.

\begin{figure}[p]
  \centering
  \begin{subfigure}[b]{0.88\linewidth}
    \includegraphics[width=\linewidth]{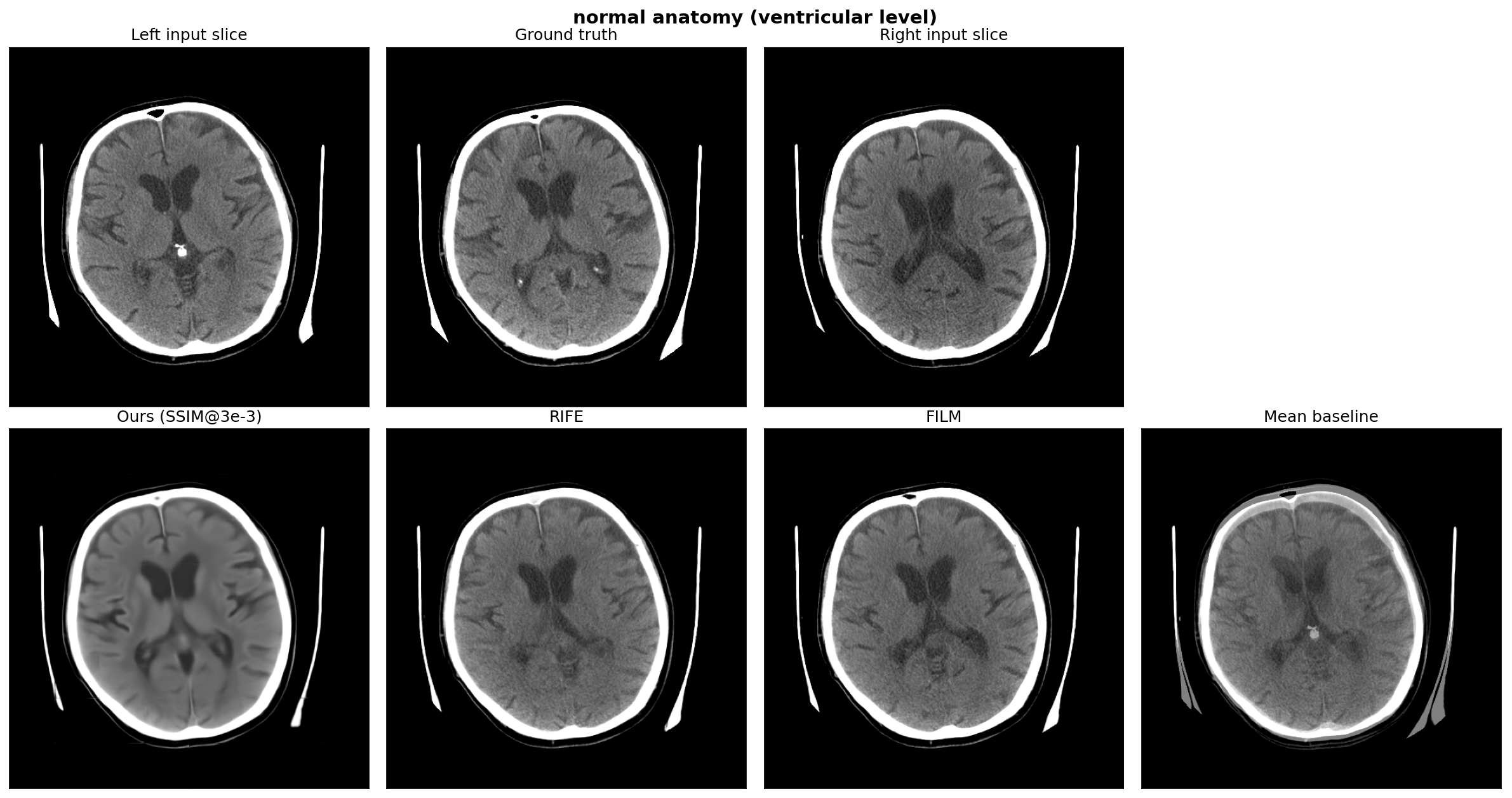}
    \caption{Patient \textsf{ID\_615f69e3}, triplet index 14 (mid-brain ventricular / basal-ganglia level; RSNA label \emph{any$=0$}); same slice as \Cref{fig:qual_main}.}
  \end{subfigure}

  \vspace{0.5em}

  \begin{subfigure}[b]{0.88\linewidth}
    \includegraphics[width=\linewidth]{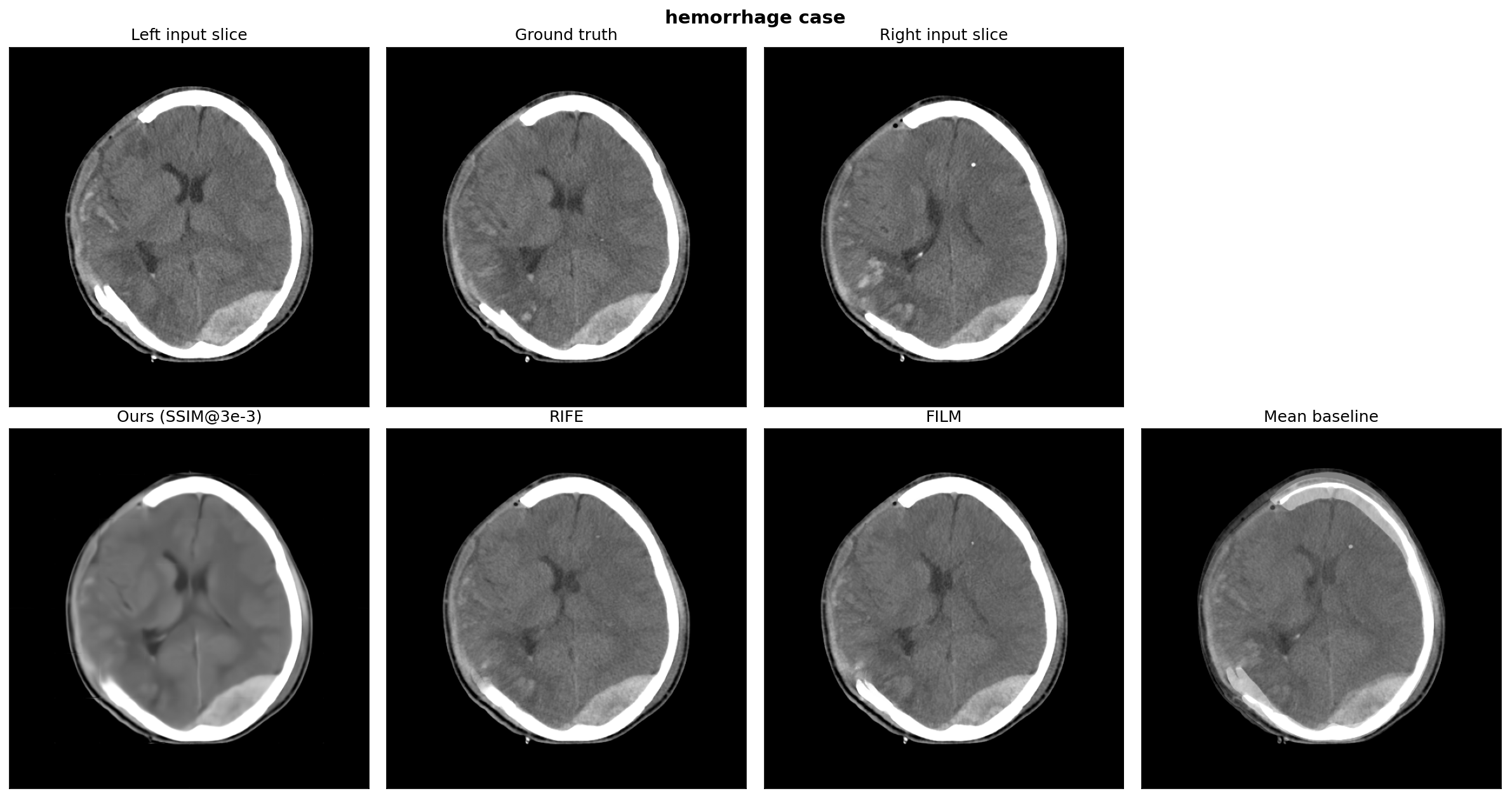}
    \caption{Patient \textsf{ID\_fc4fcd34}, triplet index 16 (mid-brain hemorrhage; positive RSNA labels for all five hemorrhage subtypes on the target slice).}
  \end{subfigure}
  \caption{Cross-method qualitative comparison on two mid-brain test slices. Top row of each panel: left input slice, ground truth, right input slice. Bottom row: our SSIM$_{\mathrm{s}}$ model, RIFE, FILM, and mean baseline. RIFE and FILM produce plausible but structurally less accurate outputs, particularly near bone boundaries and hemorrhage margins; the mean baseline is visibly blurred. Brain window; radiological convention.}
  \label{fig:vfi_comparison}
\end{figure}

\begin{figure}[p]
  \centering
  \includegraphics[width=\linewidth]{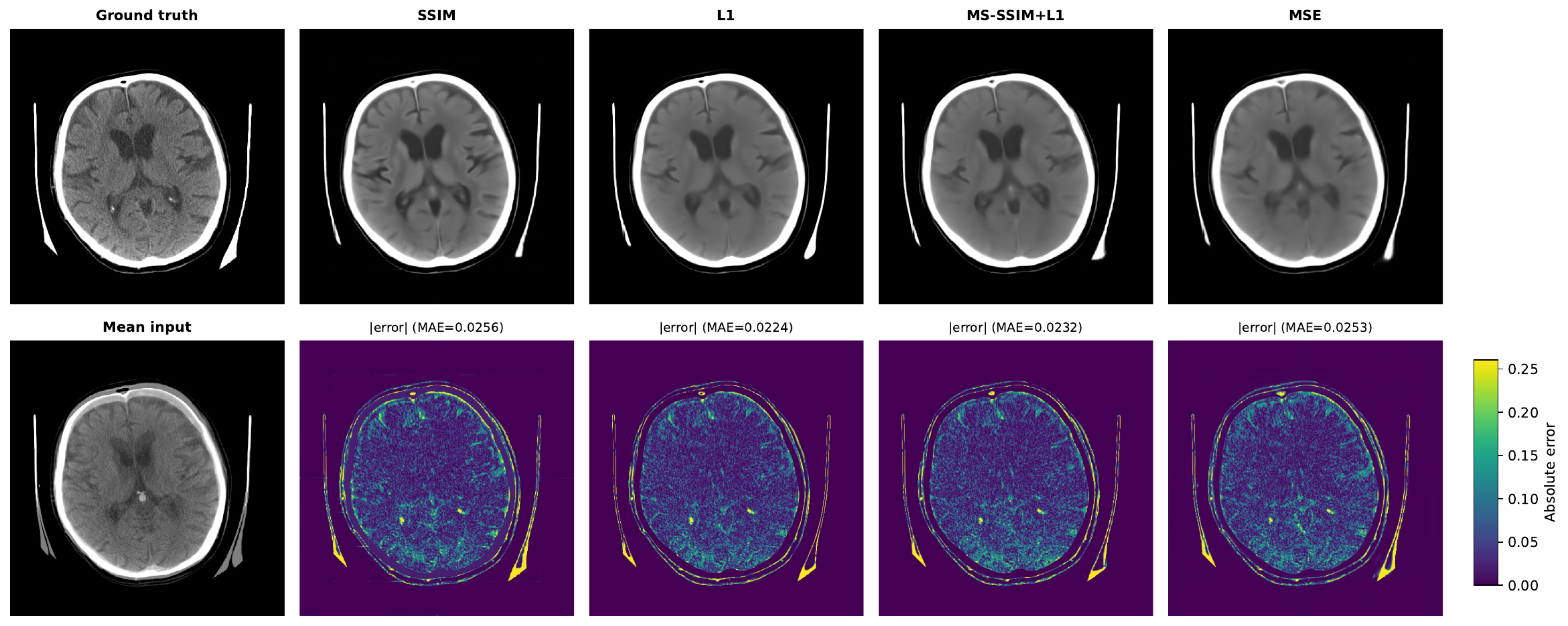}
  \caption{Predictions and pixel-wise absolute error maps for four loss functions on the same test slice as \Cref{fig:qual_main} (patient \textsf{ID\_615f69e3}, target triplet index 14). Layout is a $2\times 5$ grid. Top row: ground truth followed by model predictions for SSIM$_{\mathrm{s}}$ (lr $= 3\!\times\!10^{-3}$, batch size 32), L1, MS-SSIM+L1, and MSE (left to right). Bottom row: mean-of-adjacent-inputs baseline, followed by per-loss absolute error maps $|\hat{Y} - Y|$ in the same column order, on a shared viridis scale (colour bar at right, ``Absolute error'') whose upper bound is the $99$th percentile of the pooled error values across the four loss maps, so that the colour range is directly comparable across panels. All four losses concentrate errors at high-contrast boundaries (skull, bone/air interfaces, and the patient-support region at the posterior field of view); brain parenchyma errors are comparatively low, with a residual diffuse noise-like component. Slice-level MAE printed in each sub-panel: SSIM$_{\mathrm{s}}$ $0.0256$, L1 $0.0224$, MS-SSIM+L1 $0.0232$, MSE $0.0253$; L1 achieves the lowest MAE on this individual slice. These slice-level values are for this single example only and are not directly comparable to the patient-level means in \Cref{tab:test_patient}. Grayscale panels use the brain window (center $44$\,HU, width $128$\,HU). Radiological convention.}
  \label{fig:error_maps}
\end{figure}

\textbf{Selection rule.} From the 30 held-out test patients (968 triplets), we hand-picked mid-brain axial slices (ventricular / basal-ganglia level), where anatomy varies more smoothly than at the skull base or vertex. Selection was not random and was not driven by a ranked metric: the two patients shown are the same patients used throughout the paper for qualitative material (\Cref{fig:vfi_comparison,fig:error_maps,fig:denoising_hemorrhage,fig:denoising_normal}) to keep cross-figure consistency. These panels therefore illustrate \emph{typical} behaviour at mid-brain levels from two test cases; they are \emph{not} claimed to be worst-case or best-case, and a systematic worst-by-SSIM failure example across the full test set is not shown here (out-of-sample failure assessment is deferred to the planned radiologist reader study; see \Cref{sec:limitations}). Patient identifiers are those of the public RSNA 2019 Intracranial Hemorrhage Detection dataset~\cite{ryai2020190211}; no additional identifiers are introduced.

\textbf{Display window.} All CT images are displayed with the brain window used for preprocessing: center $44$\,HU, width $128$\,HU (range $[-20\,\text{HU}, 107\,\text{HU}]$, linearly mapped to $[0,1]$; see \Cref{sec:methods}). No auto-contrast or per-image normalisation is applied in the figures. Orientation follows radiological convention (viewer's left = patient's right).

\textbf{Pathology labels.} Hemorrhage status is taken verbatim from the RSNA slice-level labels of the target (middle) slice. \Cref{fig:qual_main} shows patient \textsf{ID\_615f69e3} (triplet index 14; target slice order 15 of 32; axial fraction $\approx 0.48$), whose RSNA label is \emph{any$=0$} on every slice; the right basal-ganglia hyperintensity is therefore not labelled as hemorrhage. The top panel of \Cref{fig:qual_hemorrhage} (and the bottom subfigure of \Cref{fig:vfi_comparison}) shows patient \textsf{ID\_fc4fcd34} (triplet index 16; axial fraction $\approx 0.52$), whose target slice carries positive RSNA labels for all five hemorrhage subtypes. The bottom panel of \Cref{fig:qual_hemorrhage} is a consecutive slice from the same \emph{normal} patient as \Cref{fig:qual_main}.

\Cref{fig:vfi_comparison} extends the comparison to include VFI baselines (RIFE and FILM) alongside our model, illustrating the structural fidelity gap that underlies the quantitative results in \Cref{tab:test_patient}: our model most closely matches the ground truth, while RIFE and FILM produce plausible but less accurate outputs, particularly near bone boundaries and hemorrhage margins.

%% file: sections/discussion.tex
\section{Discussion}
\label{sec:discussion}

\subsection{Loss--metric alignment explains the absence of a single best model}
No single loss dominates all metrics, and the differences among the top three learned models are small (\Cref{tab:test_patient,tab:paired_patient}). L1 attains the lowest MAE, the metric it directly minimizes. MS-SSIM+L1 has the numerically highest PSNR among the controlled-comparison models (L1, MS-SSIM+L1, MSE, which share optimizer settings), but the paired tests do not distinguish it from L1 or MSE on PSNR. SSIM$_{\mathrm{s}}$ ties L1 for top SSIM but pays a small MAE penalty, because SSIM's luminance and contrast terms are normalized ratios whose values are insensitive to small constant offsets that nonetheless accumulate in MAE. MSE is uniformly dominated or tied by at least one other loss on every structural metric.

These alignments are consistent with each loss excelling on the metric it most directly optimizes. However, ``best model'' is only meaningful relative to a stated evaluation criterion.

\subsection{SSIM loss fragility and the role of numerical conditioning}
\label{sec:ssim-fragility}

The most striking result is the contrast between SSIM at lr $= 3 \times 10^{-3}$ (test SSIM $0.890$) and SSIM at lr $= 8 \times 10^{-4}$ (test SSIM $0.067$, catastrophic failure). The latter's test MAE of $11.2$ (on $[0,1]$-normalized data) indicates that model outputs diverged far outside the valid intensity range rather than collapsing to a constant prediction. Both used the same architecture and data; they differed in learning rate and batch size ($32$ vs.\ $96$).

Root-cause analysis identified one dominant numerical failure mode in the SSIM stability constant $C_2 = (K_2 \cdot L)^2$: with default $K_2 = 0.03$ this yields $C_2 = 0.0009$ for data range $L = 1$, which provides inadequate numerical stabilization when covariance denominators approach zero during early training. Setting $K_2 = 0.4$ ($C_2 = 0.16$), together with the FP32 autocast exclusion of \Cref{par:ssim_stability}, eliminated the common near-zero-denominator / low-precision NaN failure at initialization. It did not eliminate residual training fragility, as shown by the divergence and degenerate-output patterns below.

Under the operational definition of \Cref{par:divergence-definition}, $18$ of the $33$ SSIM-family runs diverged. Divergent runs spanned all tested weight ratios (SSIM/L1 from 0.2/0.8 to 0.8/0.2) and learning rates ($1 \times 10^{-4}$ to $8 \times 10^{-4}$); the full per-family, per-batch-size accounting is given in \Cref{tab:divergence_accounting}. The successful SSIM configuration (batch size 32, learning rate $3 \times 10^{-3}$) avoided this failure in our registry, plausibly through larger per-sample gradient variance that acted as implicit regularization. However, because this configuration differs from the other primary models in both batch size (32 vs.\ 96) and learning rate ($3\!\times\!10^{-3}$ vs.\ $8\!\times\!10^{-4}$), the individual contributions of loss function, learning rate, and batch size cannot be disentangled. Consequently, SSIM$_{\mathrm{s}}$'s leading structural similarity should be interpreted as evidence of what SSIM loss \emph{can} achieve under favorable conditions, not as a controlled comparison against the other losses. Regardless of mechanism, the sensitivity of SSIM loss to these implementation details, where a 3.75$\times$ change in learning rate produces the difference between near-optimal test SSIM ($0.890$) and catastrophic failure ($0.067$), represents a practical hazard for practitioners adopting SSIM-based training objectives even in the presence of the $K_2$/FP32 safeguards.

\subsection{Training instability as a systematic finding}
\label{sec:training-instability}
Training divergence was concentrated in SSIM-family losses, and divergence \emph{rates} were amplified at smaller batch sizes. Under the operational definition of \Cref{par:divergence-definition}, the SSIM+L1 and MS-SSIM+L1 subfamilies accounted for the largest absolute divergence counts: $9$ of $15$ SSIM+L1 and $9$ of $14$ MS-SSIM+L1 runs diverged (see \Cref{tab:divergence_accounting}); in addition, among the $6$ SSIM+L1 runs that reached early stopping, $3$ produced degenerate validation SSIM $< 0.13$, representing a distinct failure mode not captured by the NaN-based divergence criterion. Reducing batch size from $96$ to $64$ raised the divergence rate from $10/25 = 40\%$ to $11/13 \approx 85\%$ across all registered loss families (\Cref{tab:divergence_accounting}); in particular, L1 at $\text{lr}=1\!\times\!10^{-4}$ completed at batch size $96$ but diverged at batch size $64$, illustrating that batch-size reduction can destabilise losses that were otherwise stable at the same learning rate. Among the non-SSIM losses, MSE and L1 completed training at batch size $96$ with the default learning rate ($8\!\times\!10^{-4}$).

Training divergence is rarely analyzed systematically in the literature. We argue that the divergence rate and its structure constitute a genuine finding: SSIM-family losses require explicit numerical conditioning, and batch size interacts with learning rate in ways that standard hyperparameter guidance does not address. Characterizing these divergence patterns enables other groups to avoid unstable regions of the configuration space.

\subsection{Implicit denoising as a property of regression-based synthesis}
\label{sec:noise}
A visible property of the model outputs in \Cref{fig:qual_main,fig:qual_hemorrhage} is the absence of the acquisition noise texture that is present in the original CT slices. This noise suppression is not a novel finding but a consequence of the classical regression identity from estimation theory (standard textbook result, see e.g.\ Bishop~\cite{bishop2006pattern} or Hastie \emph{et al.}~\cite{hastie2009elements}): for any square-integrable target $Y$ and input $X$,
\begin{equation}
  \arg\min_{f} \, \mathbb{E}\bigl[(f(X) - Y)^{2}\bigr] \;=\; \mathbb{E}[Y \mid X],
  \label{eq:regression_identity}
\end{equation}
the minimizer of the $L^{2}$ risk being the conditional mean. If the target decomposes as $Y = S(X) + N$ with $\mathbb{E}[N \mid X] = 0$, then $\mathbb{E}[Y \mid X] = S(X)$, so the $L^{2}$-optimal predictor recovers the conditional-mean signal even when the individual targets are noisy. This holds under square-integrability of $Y$ and the zero-conditional-mean assumption on $N$; the stronger independence $N \perp X$ simplifies variance analysis but is not required for \eqref{eq:regression_identity}. The L1 risk is minimized by the conditional median, which coincides with $S(X)$ only when the noise conditional on $X$ is symmetric about zero. SSIM-based losses are neither mean nor median estimators (the SSIM map is a non-linear, data-dependent functional of local moments), so we do not claim they inherit the identity; any smoothness in their outputs is treated as an empirical observation, not a consequence of \eqref{eq:regression_identity}.

This identity predates deep learning by decades and underpins classical regression theory. Its modern leverage in image restoration is due to Lehtinen \emph{et al.}~\cite{noise2noise}, whose Noise2Noise framework showed that when \emph{two} independent noisy observations of the same underlying signal are available as input--target pairs, a clean target is unnecessary for training: the identity still recovers $\mathbb{E}[Y \mid X] = S(X)$. Our setting is not Noise2Noise: we have a single noisy target $I_{k+1}$ conditioned on adjacent inputs $(I_k, I_{k+2})$, not two noisy observations of the same slice. What we rely on is the older and weaker conditional-expectation identity \eqref{eq:regression_identity} itself, which applies to any regression-based synthesis task trained with an $L^{2}$ (or L1, under symmetry) loss. The observed noise suppression is therefore \emph{consistent with} this identity and does not require the Noise2Noise construction.

Two caveats apply when moving from \eqref{eq:regression_identity} to calling the output ``denoised''. First, CT noise is signal-dependent (Poisson/photon counting) and spatially correlated by the reconstruction kernel~\cite{CHEN2024111355}; these properties do not invalidate \eqref{eq:regression_identity} but mean that the recovered $\mathbb{E}[Y \mid X]$ is a conditional-mean estimate under whatever joint distribution the training data induce, not a guaranteed recovery of an idealized clean signal. Second, $\mathbb{E}[N \mid X] = 0$ is plausible for reconstructed CT in approximation but cannot be verified from our data; systematic biases (beam hardening, partial-volume averaging) would appear as structured residuals rather than noise-like texture.

This implicit denoising is directly relevant to CT imaging, where noise is a well-known clinical problem. The low-dose CT reconstruction literature addresses noise using a closely related approach: training encoder--decoder networks with regression losses (typically MSE) on \emph{paired low-dose/normal-dose} CT images, where the normal-dose scan serves as a (near-)clean supervisory target. Chen \emph{et al.}~\cite{Chen2017-ml} demonstrated that a residual encoder--decoder CNN (RED-CNN) trained with MSE loss on such pairs achieved simultaneous noise suppression, structural preservation, and improved lesion detection, evaluated favorably by radiologists in a clinical reader study. The comparison to our setting is an analogy, not an equivalence: RED-CNN is fully supervised with an approximately clean target and is optimized to minimize distance to that target, whereas our network has no clean slice available and minimizes distance to the noisy acquired slice $I_{k+1}$ conditioned on its neighbors. Both setups are instances of regression-based synthesis with an $L^{2}$ or $L^{2}$-like loss and therefore share the same underlying estimator \eqref{eq:regression_identity}, but the supervisory signals and what is implicitly being estimated differ.

\begin{figure}[p]
  \centering
  \includegraphics[height=0.88\textheight,keepaspectratio]{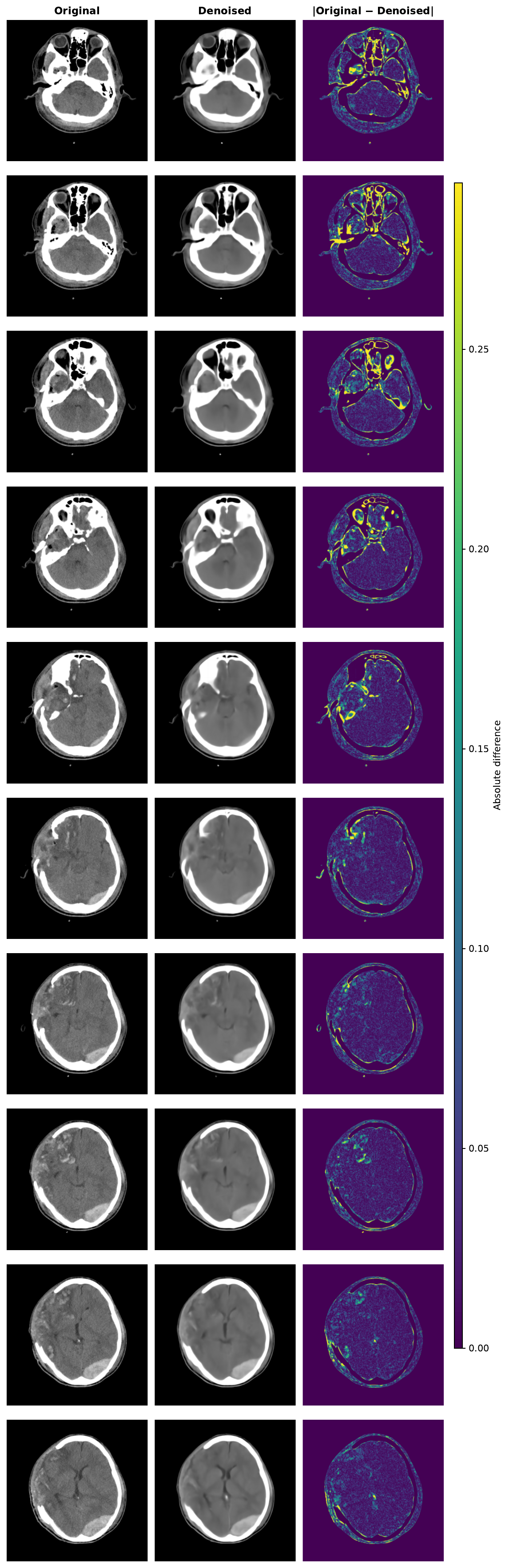}
  \caption{Implicit denoising across 10 consecutive slices (hemorrhage patient). Column~1: original noisy acquired slice. Column~2: model prediction from neighboring slices. Column~3: absolute difference, showing predominantly noise-like texture removal. Gross hemorrhagic and anatomical structures remain visible, but diagnostic preservation requires reader validation.}
  \label{fig:denoising_hemorrhage}
\end{figure}

\begin{figure}[p]
  \centering
  \includegraphics[height=0.88\textheight,keepaspectratio]{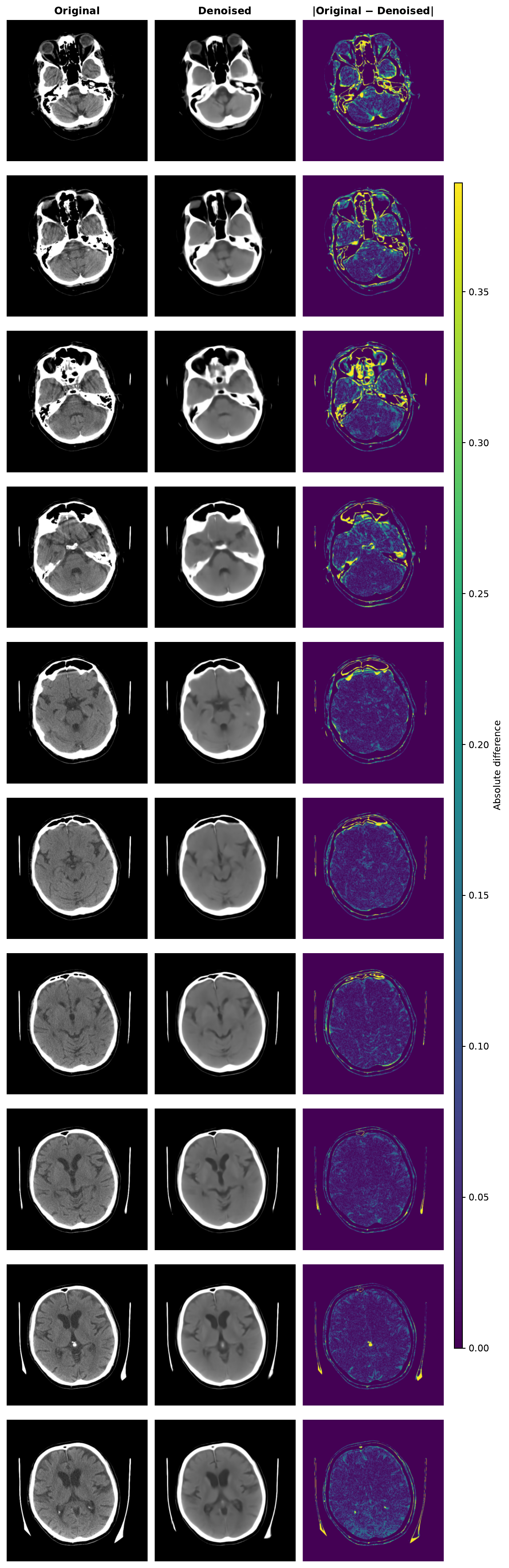}
  \caption{Implicit denoising across 10 consecutive slices at normal (non-hemorrhage) axial levels from the same patient as \Cref{fig:qual_main}. The difference maps show predominantly spatially diffuse, noise-like residuals, consistent with the pattern in \Cref{fig:denoising_hemorrhage}.}
  \label{fig:denoising_normal}
\end{figure}

\Cref{fig:denoising_hemorrhage,fig:denoising_normal} illustrate this implicit-denoising behaviour across 10 consecutive slices for a hemorrhage and a normal patient, respectively. For each acquired slice $I_{k+1}$, the model is given the two neighboring slices $(I_k, I_{k+2})$ and predicts the middle slice. The difference maps (column~3) are dominated by spatially diffuse, noise-like texture rather than obvious anatomical loss at this display scale, consistent with the model suppressing acquisition noise while retaining the main anatomical structures. The effect is visible in both pathological and normal examples; a quantitative ROI-based analysis is given in \Cref{sec:noise-results}.

From a clinical perspective, the denoised interpolated slices may improve visibility of subtle findings such as small hemorrhages or low-contrast lesions, analogous to the diagnostic benefits demonstrated for deep learning-based CT reconstruction~\cite{Chen2017-ml}. However, the altered noise texture may appear unfamiliar to radiologists accustomed to conventional CT appearance, and whether reduced noise enhances or impairs diagnostic confidence remains an open question.

A direct qualitative anchor for the noise-removal effect is the cerebrospinal fluid (CSF) in the lateral ventricles of a normal patient: CSF has approximately uniform composition and HU value, so any in-region intensity variability in the acquired slice is dominated by acquisition noise rather than tissue structure. The predicted ventricular regions in \Cref{fig:denoising_normal} are visibly smoother than the acquired counterparts while ventricular boundaries are preserved, consistent with the conditional-expectation reading of \eqref{eq:regression_identity} on a region whose underlying signal is approximately spatially constant by physiology.

The ROI-level results in \Cref{sec:noise-results} (\Cref{tab:roi_noise}) make this empirical: outputs from the pixel-wise regression losses have median in-tissue variance ${\approx}26\%$ below the noisy acquired reference, supporting the \Cref{eq:regression_identity} reading directly. The weaker denoising by SSIM$_{\mathrm{s}}$ and the variance increase in the averaging baseline are empirical comparisons rather than direct consequences of the regression identity, but they are consistent with the idea that learned regression suppresses unpredictable texture more effectively than anatomy-agnostic interpolation. Two caveats apply. First, the analysis is ROI-based and characterises within-tissue uniformity rather than absolute noise power at the image level. Second, NPS computed in uniform white-matter ROIs does not characterise the model's behaviour at high-contrast edges, where the denoising--sharpness trade-off is most clinically consequential.

\subsection{External case study: predicted residual signature on an out-of-distribution scan}
\label{sec:parity-case-study}

The RSNA-internal evidence of \Cref{sec:noise-results} shows that the trained models reduce variance in uniform white-matter ROIs, consistent with acquisition-noise suppression and with the conditional-expectation interpretation of \Cref{eq:regression_identity}. A complementary test is to take that interpretation out of the training distribution and turn it into a \emph{falsifiable prediction}: if the network is behaving as a conditional-expectation estimator and not simply memorising RSNA noise texture, its residual on a scan with a known, independently measurable structural property should follow the form the theory prescribes. We report such a test on a single out-of-distribution head CT series from a Spanish hospital that contributed no data to training. We present this as a case study, not an external-validation cohort.

\paragraph{Scan and geometry.}
The series (\textsf{HUVR-1}; acquired on a GE LightSpeed16 scanner with the SOFT reconstruction kernel; slice thickness $3.70$\,mm; in-plane spacing $0.488$\,mm; gantry tilt $\approx\!25^\circ$) contains $N=32$ reconstructed axial slices with a deliberate change of through-plane spacing along the z axis. From the DICOM \texttt{ImagePositionPatient} coordinates, the inter-slice spacing is $\approx\!2.75$\,mm for slices $k = 1\text{--}17$ (slices \emph{overlap}: nominal thickness $3.70\,\mathrm{mm} > $ spacing $2.75\,\mathrm{mm}$) and $\approx\!5.50$\,mm for slices $k = 18\text{--}32$ (slices \emph{do not overlap}: a $\approx\!1.8$\,mm gap separates adjacent $3.70$\,mm-thick slabs). This mixed-spacing geometry is incidental to the case selection but turns out to be structurally informative, as explained below.

\paragraph{Data-side structural hypothesis.}
Suppose the source reconstruction contributes a small, parity-locked additive offset to alternate axial slices on top of a smooth through-plane signal,
\begin{equation}
  \mathrm{real}_k(x, y) \;=\; s(z_k, x, y) \;+\; (-1)^k\,b(x, y),
  \label{eq:parity_model}
\end{equation}
with $|b| \ll$ tissue contrast. Such offsets arise in multi-slice helical CT reconstruction when the through-plane interpolation weights between detector rows differ systematically for odd and even reconstructed slices; the exact mechanism is scanner- and protocol-dependent and need not be assumed or verified here. Our triplet formulation predicts $\hat{I}_{k} = f(I_{k-1}, I_{k+1})$ and therefore sees inputs that share the \emph{same} parity, opposite to the target. Under \Cref{eq:parity_model}, the linear interpolation baseline $f_\mathrm{lin}(I_{k-1}, I_{k+1}) = (I_{k-1} + I_{k+1})/2$ produces the residual
\begin{equation}
  \hat{I}^{\,\mathrm{lin}}_k - \mathrm{real}_k \;=\; \underbrace{\bigl(\tfrac12 s(z_{k-1}) + \tfrac12 s(z_{k+1}) - s(z_k)\bigr)}_{\text{smooth anatomy residual}} \;-\; 2\,(-1)^k\,b,
  \label{eq:linear_residual}
\end{equation}
so \emph{any} data-side parity bias appears in the residual with \emph{twice} its amplitude and opposite sign on alternate slices. This algebraic doubling makes the linear baseline a built-in amplifier for data-side parity structure, and provides a per-slice scalar test: take the per-slice mean of the residual and look at the Nyquist-frequency (period-two) bin of its DFT along the slice axis.

If the parity-locked component is not inferable from $(I_{k-1}, I_{k+1})$ beyond its shared input parity, a learned conditional-expectation estimator should regress that target-specific component towards zero. Under \Cref{eq:regression_identity}, the residual of a trained model should therefore retain the parity structure but at \emph{reduced} amplitude relative to the linear baseline. The conditional-expectation interpretation predicts two things simultaneously on this scan: (i) a period-two component in the linear-baseline residual with Nyquist amplitude $\approx 2b$; and (ii) systematic attenuation of that amplitude by trained models.

\begin{figure}[t]
  \centering
  \includegraphics[width=\linewidth,keepaspectratio]{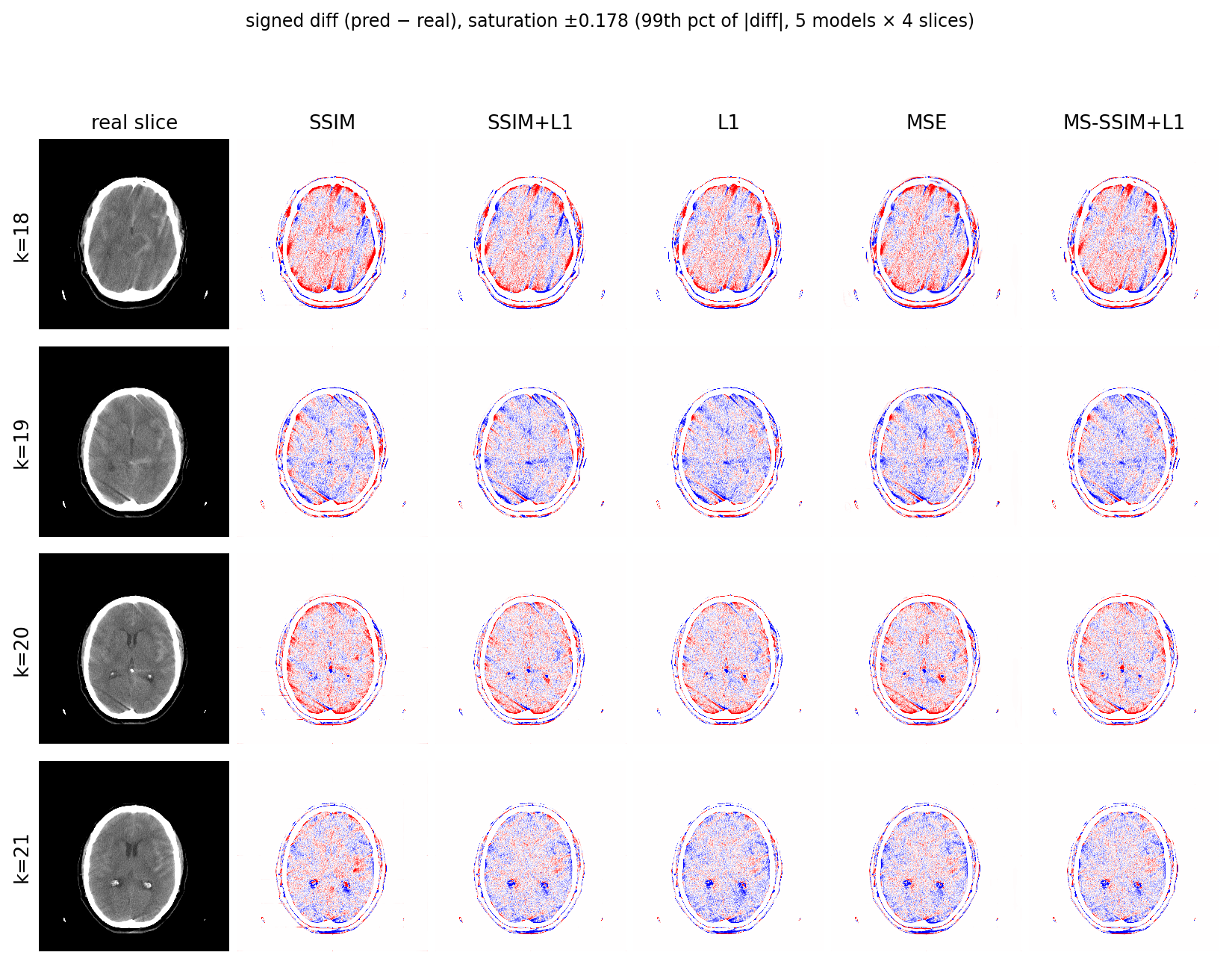}
  \caption{Four consecutive reconstructed slices from \textsf{HUVR-1} in the coarse-spacing region ($k \geq 18$; $\approx\!1.8$\,mm gap between non-overlapping $3.70$\,mm-thick slabs) and their model residuals. Leftmost column: acquired slice (brain window, center $44$\,HU, width $128$\,HU). Remaining columns: signed diff $\hat{I}_k - \mathrm{real}_k$ for each trained loss; red$>0$, blue$<0$; saturation $\pm$ the 99th-percentile absolute diff across all models and coarse-region slices, so colour magnitudes are directly comparable across rows and across losses. The sign of the diff flips between consecutive slices and is consistent across every loss, the signature predicted by \Cref{eq:linear_residual}. Quantitative evidence and spectral decomposition are given in \Cref{fig:parity_quant} and \Cref{tab:parity_amplitudes}. Images shown in radiological convention.}
  \label{fig:parity_diff_strip}
\end{figure}

\begin{figure}[t]
  \centering
  \includegraphics[width=0.92\linewidth,keepaspectratio]{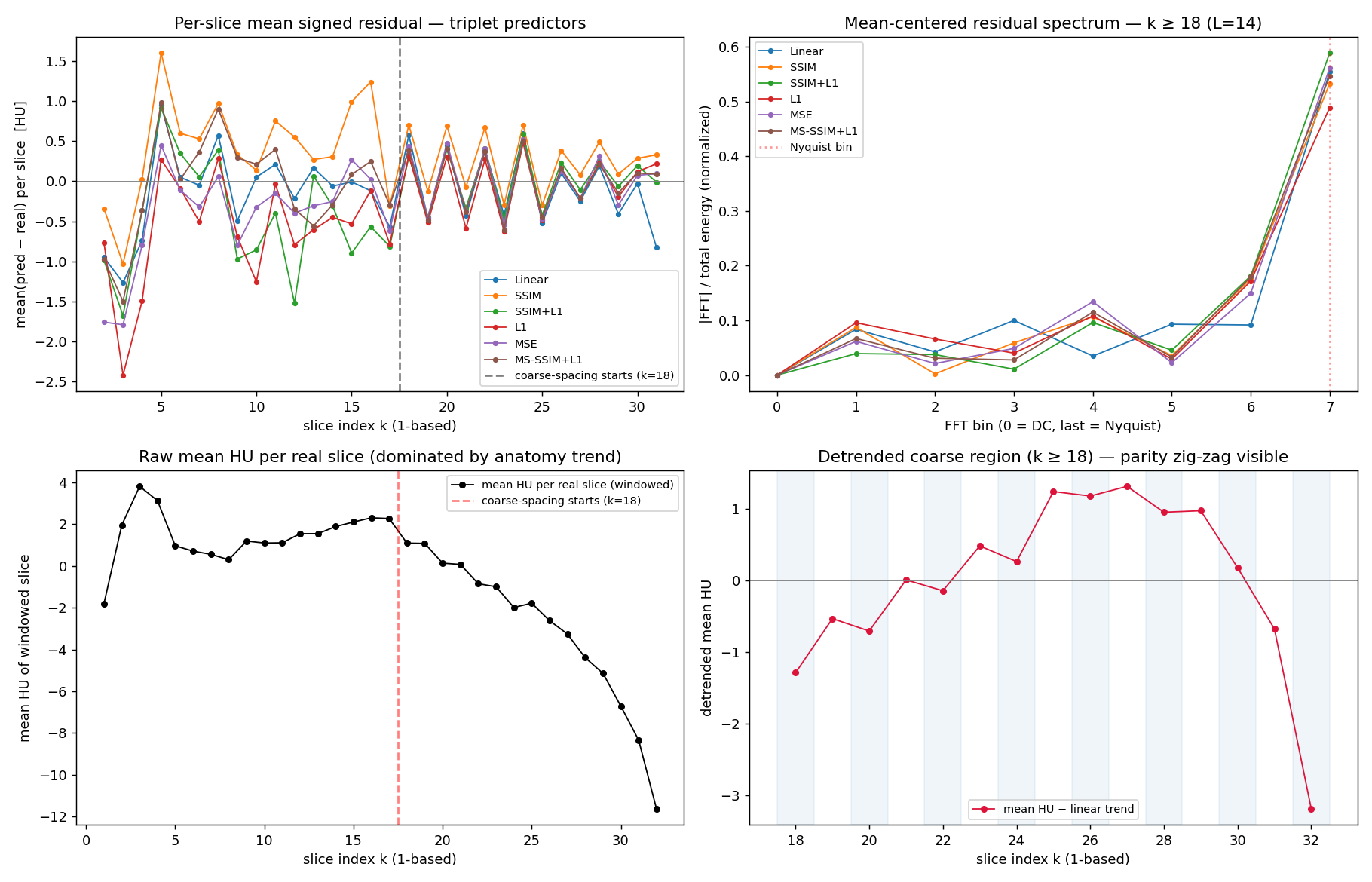}
  \caption{Quantitative decomposition of the residual on \textsf{HUVR-1}. Top-left: per-slice mean signed residual in HU for each predictor, across the middle-slice range $k \in [2, N{-}1]$; the dashed vertical line marks the transition from the overlap region to the coarse-spacing region. Top-right: normalised spectrum $|\mathrm{FFT}|/\text{total}$ of the mean-centred residual restricted to the coarse-spacing region ($k \geq 18$, $L = 14$), with the rightmost (Nyquist) bin corresponding to period-two structure along the slice axis. Bottom-left: mean HU value of each brain-windowed acquired slice (no predictor involved); the dashed vertical line marks the spacing regime change. Bottom-right: the same real-data signal in the coarse region after subtracting a linear trend, revealing a period-two oscillation; alternate slices are shaded to make the parity alignment explicit. The Nyquist amplitude of this bottom-right signal, $0.22$\,HU (unrounded $0.217$\,HU), is the direct measurement of $b$ in \Cref{eq:parity_model}. The algebraic prediction $2b \approx 0.43$\,HU matches the measured linear-baseline residual amplitude ($0.40$\,HU) to within $0.03$\,HU.}
  \label{fig:parity_quant}
\end{figure}

\paragraph{Empirical test.}
\Cref{fig:parity_diff_strip} shows diff maps $\hat{I}_k - \mathrm{real}_k$ for four consecutive slices in the coarse-spacing region across the real reference and all five trained losses; the red/blue sign flip on alternate slices is directly visible and is consistent across every loss. We measured the data-side amplitude $b$ directly, with no predictor involved, by taking the mean HU value of each brain-windowed real slice and reading off the Nyquist component of its DFT along the slice axis (\Cref{fig:parity_quant}, bottom-right): in the coarse region $k \geq 18$ the parity amplitude is $0.22$\,HU. The linear-baseline residual Nyquist amplitude, measured the same way on the same slice subset, is $0.40$\,HU, within $0.03$\,HU of the $2b \approx 0.43$\,HU predicted by \Cref{eq:linear_residual} (using unrounded $b = 0.217$\,HU; the rounded display value $2 \times 0.22 = 0.44$\,HU differs from the measurement by $0.04$\,HU and both match the prediction to within display precision). For the five trained losses the residual parity amplitude is $0.30$--$0.33$\,HU (mean $0.31$\,HU), a $\approx\!22\%$ reduction relative to the linear baseline. \Cref{tab:parity_amplitudes} summarises the amplitudes and the fraction of residual energy at the Nyquist bin; in the coarse region the period-two component accounts for $49$--$59\%$ of the mean-centred residual energy for every predictor, i.e.\ dominates the spectrum.

\paragraph{Interpretation.}
Three observations support the conditional-expectation reading. First, the linear-baseline amplitude matches the theoretical $2b$ doubling to well below $1$\,HU, confirming that the period-two component is a property of the DICOM pixel data rather than a model fabrication. Second, every trained loss attenuates this data-side amplitude (\Cref{tab:parity_amplitudes}) without being given an explicit parity indicator: this is consistent with a regressor treating target-specific, poorly predictable components as noise and pushing them towards zero, the same general mechanism invoked for the ROI-level noise suppression of \Cref{sec:noise-results} but tested on a different signal (a structural parity bias, not stochastic acquisition noise) on out-of-distribution data. Third, the effect is clearest in the coarse-spacing region: the full-range parity energy fraction is $13\%$ for the linear baseline and similar for every trained model, versus $49$--$59\%$ in the coarse subregion. A plausible explanation is that overlapping slices below $k=18$ share more z-content ($\approx\!26\%$ z-overlap: thickness $3.70$\,mm vs.\ spacing $2.75$\,mm), increasing the smooth-anatomy residual in \Cref{eq:linear_residual} and making the period-two component less dominant. The geometry of this particular scan therefore provides a cleaner region in which to observe the parity residual.

A residual with this structure is not a failure of the model. Any part of the target that is not inferrable from the inputs must, by construction, leave a trace in the residual; the absence of such a trace would indicate leakage or memorisation. The result is therefore better read as: on a hospital scan the models have never seen, with a measurable data-side property whose interaction with the triplet formulation is derivable from first principles, each trained loss produces a residual of the shape and magnitude the conditional-expectation identity predicts.

\paragraph{Scope.}
Three limitations apply to the quantitative claims. (i) $n = 1$ scan from one scanner vendor and one protocol; the $\approx\!22\%$ attenuation number should not be read as a population statistic and in particular cannot be generalised to other scanners or reconstruction kernels. (ii) Amplitudes are measured on brain-windowed pixel data ($[-20\,\text{HU}, 107\,\text{HU}]$), so values outside that window are clipped before the residual is formed; for the slice-mean-level signal this is negligible but the measurement cannot speak to parity effects in bone or extracranial air. (iii) The mechanism attribution in \Cref{eq:parity_model} is phenomenological: the parity bias is demonstrated in the pixel data, but the GE reconstruction pipeline that produced it is not observed by us. The linear-baseline doubling test and the model-side attenuation are independent of any particular physical cause.

\begin{table}[t]
  \centering
  \caption{Parity amplitude of the per-slice mean residual on \textsf{HUVR-1}, along with the direct measurement of the data-side amplitude $b$ from the acquired real series. \textbf{Parity frac.} is the fraction of mean-centred residual energy at the Nyquist (period-two) bin of the slice-axis DFT. \textbf{parity amp (coarse)} is the Nyquist amplitude in HU on the coarse-spacing subrange ($k \geq 18$, $L = 14$). The direct real-series row reports $b$ itself, not a residual. Theory predicts a linear-baseline residual amplitude of $2b \approx 0.43$\,HU (using the unrounded measurement $b = 0.217$\,HU); measured is $\mathbf{0.40}$\,HU (bolded in the table).}
  \label{tab:parity_amplitudes}
  \small
  \begin{tabular}{lrrr}
  \toprule
  Predictor                & Parity frac. (full) & Parity frac. (coarse) & Parity amp (coarse, HU) \\
  \midrule
  Linear baseline          & $0.13$ & $0.55$ & $\mathbf{0.40}$ \\
  SSIM                     & $0.11$ & $0.53$ & $0.30$ \\
  SSIM+L1                  & $0.08$ & $0.59$ & $0.30$ \\
  L1                       & $0.10$ & $0.49$ & $0.31$ \\
  MSE                      & $0.10$ & $0.56$ & $0.33$ \\
  MS-SSIM+L1               & $0.10$ & $0.55$ & $0.31$ \\
  \midrule
  Real series (direct)     & $0.01$ & $0.05$ & $0.22$ \\
  \bottomrule
  \end{tabular}
\end{table}

\subsection{Effect of pathology on interpolation difficulty}
\label{sec:loss-pathology}
The hemorrhage-vs-normal SSIM gap is consistent across every evaluated method and survives the patient-level cluster bootstrap in 7 of 8 cases (Mean baseline marginal at $p_\mathrm{cb}=0.056$; \Cref{tab:hemorrhage_strat}; full inferential tests in \Cref{sec:appendix_hemorrhage_tests}, \Cref{tab:hemorrhage_tests}). For the classical and VFI baselines, MAE and PSNR nominally \emph{favour} hemorrhage slices, an artefact of globally blurred outputs whose error is dominated by parenchymal texture rather than focal lesion structure. For the learned models, by contrast, the small nominal MAE and PSNR gaps do not survive the patient-level bootstrap, so the pathology effect in that regime is an SSIM phenomenon rather than a uniform drop in reconstruction accuracy. This is consistent with SSIM penalising the structural ambiguity introduced by focal, high-contrast lesions (localised error that per-pixel summaries dilute over the whole slice), and the related ill-posedness when a lesion lies between acquired slices is discussed under partial-volume effects (\Cref{sec:limitations}).

\subsection{Limitations}
\label{sec:limitations}

\textbf{Single anatomy, acquisition protocol, and operating range.}
All data are axial head CT from the RSNA 2019 dataset, acquired and preprocessed within the brain window (center $44$\,HU, width $128$\,HU; $[-20\,\text{HU}, 107\,\text{HU}]$). Preprocessing saturates values outside this range, so the model neither interpolates nor denoises bone, calcification, or extracranial air in their native HU ranges: those structures enter the network already clipped to the window boundaries. This is a scope choice matched to the diagnostic use case (parenchymal hemorrhage assessment) rather than a general-purpose CT interpolator. Generalization to other anatomies (chest, abdomen), other modalities (MRI), or other clinical windows (bone, lung, soft-tissue) is untested and would require retraining on the corresponding windowed data; head CT also has relatively homogeneous tissue contrast compared to abdominal CT, which may favor simpler losses.

\textbf{Metrics against a noisy reference, and absence of perceptual/distribution-level metrics.}
All reported MAE/SSIM/PSNR compare the prediction to the acquired middle slice, which itself contains acquisition noise. A model that perfectly recovered the conditional-expectation signal (\Cref{sec:noise}) would differ from this reference precisely by the noise it suppresses, and would therefore score somewhat worse on per-pixel metrics than a model that reproduces the noise. Small metric differences between top-performing losses should thus be read as potentially reflecting noise-matching fidelity rather than signal-recovery quality, which motivates the quantitative denoising analysis in \Cref{sec:noise-results}. The reported metrics are also all per-image structural or intensity-based: learned perceptual metrics (e.g., LPIPS) and distribution-level metrics (e.g., FID) are not included as reported outcomes because their calibration for medical imaging is not established and they rely on feature extractors trained on natural images; these could nonetheless reveal failure modes such as mode collapse or systematic texture shifts that per-image structural metrics miss.

\textbf{Partial-volume effects.}
At the typical inter-slice spacing of ${\sim}5$\,mm, structures smaller than the slice thickness (e.g., small vessels, thin hemorrhage layers) may appear on one acquired slice and be absent from its neighbors. The model is then asked to interpolate anatomy not represented in either input, which is fundamentally ill-posed and may produce hallucinated or missing structures in the synthesized slice. The same ill-posedness affects head-support and head-holder material whose cross-sectional footprint differs between $I_{k}$ and $I_{k+2}$: the network tends to output a diffuse low-intensity halo at those locations rather than either of the two input footprints, and this is the source of the posterior-right haze visible on the prediction panels of \Cref{fig:qual_main,fig:qual_hemorrhage}. The artefact is extracranial and does not affect parenchymal structures, but it is a visible reminder that the model minimises a regression loss under through-plane ambiguity rather than performing geometry-aware reasoning about the support device.

\textbf{No clinical validation and limited cross-architecture comparison.}
Interpolated slices have not been evaluated by radiologists for diagnostic acceptability; metric improvements do not necessarily translate to clinical utility, and subtle artifacts in synthesized slices could be diagnostically misleading. On the architectural side, we benchmark against pretrained VFI methods (RIFE~\cite{RIFE_ECCV2022}, FILM~\cite{reda2022film}) as external baselines, but do not compare against dedicated medical image interpolation architectures such as SAINT~\cite{SAINT} or I$^3$Net~\cite{Song2024INetII}, nor against alternative encoder backbones. By design, this work holds architecture constant (U-Net + EfficientNetV2-S) to isolate the effect of loss function choice; a blinded radiologist reader study and a systematic cross-architecture comparison under matched training budget are identified as the highest-priority next steps in \Cref{sec:future-work}.

\subsection{Future work}
\label{sec:future-work}
Several directions are prioritized.
\begin{itemize}[leftmargin=*, itemsep=1pt]
  \item \textbf{Radiologist reader study.} Conduct a blinded reader study with $N{=}3$ board-certified radiologists using a forced-choice (or Likert) diagnostic-acceptability rating on acquired vs.\ synthesized slices, with diagnostic confidence and acceptability as the primary endpoint and inter-rater agreement (Cohen's/Fleiss' $\kappa$) as a secondary endpoint, closing the clinical-validation gap flagged in \Cref{sec:limitations}; a natural downstream extension is to train an intracranial hemorrhage classifier of fixed architecture on native-spacing vs.\ interpolated-spacing volumes, building on prior detection work on this dataset~\cite{jimaging9020037}, to test whether the improved through-plane resolution and implicit denoising translate into higher detection accuracy.
  \item \textbf{Cross-anatomy validation.} Re-evaluate the loss comparison and denoising findings on non-head CT with different HU windows and tissue contrasts, e.g.\ public chest and abdominal CT datasets, to establish whether the present recommendations generalize or are anatomy-specific.
  \item \textbf{Cross-architecture comparison.} Benchmark the U-Net + EfficientNetV2-S backbone against dedicated medical slice-synthesis architectures (SAINT~\cite{SAINT} and I$^3$Net~\cite{Song2024INetII}) under a matched training budget and with loss held constant at MS-SSIM+L1, isolating architecture effects from loss-function and optimization effects.
\end{itemize}

%% file: sections/conclusion.tex
\section{Conclusion}
\label{sec:conclusion}
We have presented a deep learning system for head CT that halves the effective through-plane spacing while simultaneously producing denoised outputs, delivering interpolation and implicit denoising from a single pass.

The system, based on a U-Net with EfficientNetV2-S encoder, outperforms classical interpolation and pretrained video frame interpolation baselines on all structural metrics. A systematic evaluation of $41$ configurations reveals that no single loss dominates all metrics: L1 produces the lowest pixel error, MS-SSIM+L1 yields the highest PSNR, and SSIM$_{\mathrm{s}}$ and L1 tie for the highest structural similarity. The inter-loss differences among the top-performing models (SSIM$_{\mathrm{s}}$, L1, MS-SSIM+L1) are mostly significant in patient-level paired testing ($p < 10^{-8}$ for most comparisons) but small in magnitude ($\Delta$SSIM $\leq 0.002$). At the controlled hyperparameter setting (lr $= 8\!\times\!10^{-4}$, batch size 96), MS-SSIM+L1 emerges as a reliable default, achieving competitive performance across all metrics; the family as a whole remains stability-sensitive at smaller batch sizes (\Cref{tab:divergence_accounting}).

A key finding from the training process is the systematic instability of SSIM-family losses, traced to insufficient denominator conditioning ($K_2 = 0.03$): SSIM-containing configurations accounted for $18/22$ divergent runs under the operational definition of \Cref{par:divergence-definition}, and increasing $K_2$ to $0.4$ eliminates the dominant failure mode. Residual divergence at smaller batch sizes persists even with this safeguard, so $K_2 = 0.4$ should be treated as necessary but not sufficient. This cautions against treating SSIM-based objectives as drop-in replacements for pixel-wise losses.

Consistent with the classical conditional-expectation property of regression losses, leveraged in the Noise2Noise framework by Lehtinen \emph{et al.}~\cite{noise2noise}, the regression-trained models produce inherently denoised interpolations: the network learns the anatomical signal rather than the acquisition noise, a property well characterized in the low-dose CT literature~\cite{Chen2017-ml}. This connects slice interpolation to the established benefits of deep learning-based CT noise reduction.

Key limitations include the single-anatomy scope (head CT only) and the absence of clinical validation; cross-anatomy generalization and a radiologist reader study to assess the diagnostic impact of the denoised interpolations are planned as next steps before clinical deployment.

%% file: sections/acknowledgements.tex
\section*{Acknowledgements}
MAGN thanks IMUS-Maria de Maeztu grant CEX2024-001517-M (Apoyo a Unidades de Excelencia Mar\'ia de Maeztu) for supporting this research, funded by MICIU/AEI/10.13039/501100011033. The HUVR patient whose scan contributed to this study was recruited thanks to a grant from Consejer\'ia de Igualdad, Salud y Pol\'iticas Sociales de Andaluc\'ia, Spain (PI-0136-2012).

\appendix
\sloppy

%% file: sections/data_availability.tex
\section{Data Availability and Reproducibility}
\label{sec:appendix_repro}
\textbf{Data.} The RSNA 2019 Intracranial Hemorrhage Detection dataset~\cite{ryai2020190211} is publicly available through the Kaggle competition platform under the competition's data-use agreement; redistribution of the DICOM series is therefore not permitted, and researchers must accept the agreement on Kaggle to download the raw data. The preprocessing pipeline that converts the raw archive to the windowed PNG tensors used here is released with the paper, configured through the \texttt{DATASETS\_DIR} environment variable.

\textbf{External case-study scan.} The out-of-distribution scan analysed in \Cref{sec:parity-case-study} (\textsf{HUVR-1}) was obtained from Hospital Universitario Virgen del Roc\'io, Seville, Spain. The research project was overseen and approved by the Ethics Committee of the Hospital Universitario Virgen del Roc\'io (Cod.\ CEI2012PI/228), and patient identifiers were removed prior to transfer. The scan is used exclusively for illustration of the residual-structure argument of \Cref{sec:parity-case-study}; no training or hyperparameter selection used data from this source. Redistribution of the raw DICOM series is not permitted under the data-use agreement. Derived artefacts (anonymised per-slice mean HU values and per-predictor residual Nyquist amplitudes reported in \Cref{tab:parity_amplitudes}) are released with the paper so that the case study can be re-inspected without access to pixel data.

\textbf{Splits.} The train/validation/test patient assignment is materialized as a \texttt{split} column in the preprocessed dataset table (30 test patients, stratified by hemorrhage subtype; train/val 80/20 of the remaining patients). Dataset-side ordering and augmentation sampling use \texttt{numpy.random.default\_rng(seed=42)}; the same \texttt{seed}$=42$ is used for the statistical bootstrap throughout. Consistent with \Cref{sec:training}, we retain \texttt{cudnn.benchmark=True} and do not enforce kernel-level determinism, so bit-exact reproducibility across GPUs or driver versions is not guaranteed.

\textbf{Environment.} Python $3.13$, PyTorch $2.10.0$ with bundled CUDA~$12$ runtime (\texttt{nvidia-cudnn-cu12} $9.10.2$), torchvision $0.25.0$, \texttt{timm} $1.0.24$ (EfficientNetV2-S encoder via \texttt{segmentation-models-pytorch} $0.5.0$), \texttt{pytorch\_msssim} $1.0.0$, and \texttt{lpips} $0.1.4$. Exact pins for all transitive dependencies are frozen in the released lockfile and declared in the project manifest; \texttt{uv sync} reconstructs the environment. Training uses PyTorch automatic mixed precision with \texttt{torch.compile} (default mode); the SSIM/MS-SSIM \texttt{float32} carve-out is described in intervention~2 of \Cref{sec:losses}.

\textbf{Hardware and compute.} All training runs were executed on a single NVIDIA GeForce RTX~3080~Ti ($12$\,GB VRAM; driver $535.247.01$). Per-epoch wall-clock is $\approx\!400$--$550$\,s at the default batch size of $96$; individual runs terminated by early stopping (patience~$15$) at $20$--$40$ epochs, corresponding to $\approx\!2$--$6$\,h each. The full 41-configuration registry therefore represents on the order of $120$\,GPU-hours. Per-run configuration and per-epoch wall-clock are released alongside the paper.

\textbf{Artefacts.} All tables and figures derive from tabular artefacts released alongside the paper, covering patient- and slice-level metrics and their aggregated summaries, paired Wilcoxon tests with BH-adjusted $q$-values, hemorrhage stratification, VFI baseline metrics, DICOM spacing characterization, and model complexity. These are regenerated end-to-end from trained checkpoints by the released analysis pipeline.

\textbf{Entry points.} After environment setup and dataset-path configuration, the training pipeline of \Cref{sec:training} is reproduced through the entry-point commands documented in the repository README: queueing the 41 configurations, training, and inspecting metrics. Source, configurations, and tabular artefacts are released at \url{https://github.com/Keredu/deep-slice-interpolation}.

%% file: sections/experiment_scope.tex
\section{Experiment Scope}
\label{sec:appendix_registry}
The configuration registry contains $41$ queued configurations spanning the six loss families evaluated in the primary paper (\Cref{sec:losses}): MSE, L1, SSIM, MS-SSIM, SSIM+L1, and MS-SSIM+L1.

\paragraph{Operational definition of divergence.}
\label{par:divergence-definition}
Throughout this paper, a training run is called \emph{divergent} if and only if training terminated prematurely because a NaN or $\pm\infty$ value was detected in the loss (registry status \texttt{NAN\_VALUE\_DETECTED}). A run that reached the early-stopping criterion is called \emph{completed} (status \texttt{EARLY\_STOPPING}); this includes a small number of runs that completed with degenerate outputs (discussed in \Cref{sec:ssim-fragility}), which are flagged separately when relevant and are \emph{not} counted as divergent under this definition. The definition is machine-checked from the per-run termination record rather than derived from validation-curve heuristics, so every count below is directly reproducible from the released training registry.

Under this definition, $22$ of the $41$ runs diverged and $19$ completed to early stopping. \Cref{tab:divergence_accounting} gives the full accounting, cross-tabulated by loss family and batch size; this is the single authoritative source for the counts cited in \Cref{sec:ssim-fragility} and in the divergence discussion of \Cref{sec:discussion}.

\begin{table}[!htbp]
  \centering
  \small
  \begin{tabular}{l r r r r r r}
    \toprule
    \textbf{Loss family} & \textbf{bs=32} & \textbf{bs=64} & \textbf{bs=96} & \textbf{Total} & \textbf{Diverged} & \textbf{Completed} \\
    \midrule
    \multicolumn{7}{l}{\textit{SSIM-containing}} \\
    SSIM              & 2/0 & 0/0 & 1/0 & 3  & 0  & 3 \\
    MS-SSIM           & 0/0 & 0/0 & 1/0 & 1  & 0  & 1 \\
    SSIM+L1           & 0/1 & 1/1 & 5/7 & 15 & 9  & 6 \\
    MS-SSIM+L1        & 0/0 & 0/8 & 5/1 & 14 & 9  & 5 \\
    \midrule
    \textbf{SSIM-family subtotal} & 2/1 & 1/9 & 12/8 & 33 & 18 & 15 \\
    \midrule
    \multicolumn{7}{l}{\textit{Non-SSIM}} \\
    MSE               & 0/0 & 1/0 & 1/1 & 3  & 1  & 2 \\
    L1                & 0/0 & 0/2 & 2/1 & 5  & 3  & 2 \\
    \midrule
    \textbf{Non-SSIM subtotal}    & 0/0 & 1/2 & 3/2  & 8  & 4  & 4 \\
    \midrule
    \textbf{Grand total}          & 2/1 & 2/11 & 15/10 & 41 & 22 & 19 \\
    \bottomrule
  \end{tabular}
  \caption{Per-family, per-batch-size accounting of the training registry. Each cell shows \emph{completed\,/\,diverged} at that batch size. The ``Total'' column sums completed + diverged for that family; ``Diverged'' uses the operational definition in \Cref{par:divergence-definition} (registry status \texttt{NAN\_VALUE\_DETECTED}). Counts are reproducible from the released training registry.}
  \label{tab:divergence_accounting}
\end{table}

SSIM-containing losses account for $18/22$ ($\approx\!82\%$) of divergent runs; within that subtotal, divergence is concentrated in the two-term combinations SSIM+L1 ($9/15$) and MS-SSIM+L1 ($9/14$). Among the $4$ divergent non-SSIM runs, $2$ occurred at batch size $96$ and $2$ at batch size $64$. The \emph{rate} of divergence is highest at batch size $64$ ($11/13 \approx 85\%$), lower at batch size $96$ ($10/25 = 40\%$), and lowest at batch size $32$ ($1/3$), indicating that reducing batch size amplified per-run divergence prevalence. The complete experiment registry, including per-run learning rates, seeds, and termination epochs, is released with the source repository.

%% file: sections/appendix_hemorrhage_tests.tex
\section{Full hemorrhage-stratification inferential tests}
\label{sec:appendix_hemorrhage_tests}
This appendix gives the full inferential table summarised in \Cref{sec:hemorrhage-strat} (\Cref{tab:hemorrhage_strat}). For each of the 8 methods and 3 primary metrics (SSIM, MAE, PSNR), we ran a two-sided Mann--Whitney $U$ test%
\footnote{\cite{mannwhitney1947}.
We use \texttt{scipy.stats.mannwhitneyu} (two-sided, continuity corrected, \texttt{method='auto'}). The two groups are different slices, so the paired Wilcoxon of \Cref{sec:stats} does not apply and the two-sample rank-sum test is the appropriate nonparametric choice.}
comparing hemorrhage- and normal-slice metric distributions, and applied Benjamini--Hochberg FDR control~\cite{benjamini1995controlling} \emph{jointly over the full $8\times 3 = 24$-test family} to obtain $q$-values. The analysis was designed after seeing the descriptive results of \Cref{tab:hemorrhage_strat} and is therefore \emph{post-hoc / exploratory}; it was not pre-registered.

Because slices within a patient are not i.i.d., the asymptotic Mann--Whitney $p$ overstates evidence. We therefore also report a cluster-bootstrap $p$ ($p_\mathrm{cb}$): resample patients with replacement (10{,}000 resamples), recompute the standardized $U$-statistic $z = (U - n_1 n_2 / 2)/\sqrt{n_1 n_2 (n_1 + n_2 + 1)/12}$ on each resample, and invert the two-sided percentile CI for $z$ at zero~\cite{Efron1979}. We treat $p_\mathrm{cb}$ as the primary independence-aware $p$-value; the BH-corrected asymptotic $q$ is reported for transparency but is not corrected for within-patient clustering. Effect size is reported as (i) the Hodges--Lehmann shift estimator%
\footnote{\cite{hodgeslehmann1963}.
Median of all pairwise differences $x^\mathrm{hem}_i - x^\mathrm{norm}_j$, with 95\% cluster-bootstrap CI from the same 10{,}000 patient resamples.}
and (ii) the rank-biserial correlation%
\footnote{\cite{kerby2014simple}.
$r = 2U_1/(n_1 n_2) - 1$ with $U_1 = U(\text{hem}, \text{norm})$; positive $r$ means hemorrhage slices score higher (i.e.\ better for SSIM/PSNR, worse for MAE).}
Full numerical results for the three primary metrics in this exploratory analysis are released with the source repository.

\begin{table}[!htbp]
  \centering
  \caption{Exploratory (post-hoc) inferential tests for \Cref{tab:hemorrhage_strat}: Mann--Whitney $U$ comparing hemorrhage ($n_\mathrm{h}=396$) vs.\ normal ($n_\mathrm{n}=572$) slice-level metric distributions, two-sided. Shift = Hodges--Lehmann $\widehat\Delta$ with 95\% cluster-bootstrap CI (brackets). $r$ = rank-biserial effect size; sign convention: positive $r$ = hemorrhage scores higher (i.e.\ better for SSIM/PSNR, worse for MAE). $p_\mathrm{asym}$ = asymptotic Mann--Whitney $p$; $p_\mathrm{cb}$ = patient-level cluster-bootstrap $p$ (10{,}000 resamples); $q_\mathrm{BH}$ = Benjamini--Hochberg adjusted $q$ from the asymptotic $p$, computed jointly over the full $8\times 3 = 24$-test family. The Decision column summarises the joint outcome: $\checkmark$ = BH-significant and bootstrap-robust ($q_\mathrm{BH} < 0.05$ and $p_\mathrm{cb} < 0.05$); $\dagger$ = BH-significant but not bootstrap-robust ($q_\mathrm{BH} < 0.05$ and $p_\mathrm{cb} \geq 0.05$), i.e.\ the asymptotic BH decision does not survive the patient-level bootstrap; \textemdash\ = not BH-significant ($q_\mathrm{BH} \geq 0.05$). Recall that $q$-values do not correct for within-patient clustering, which is why we treat $p_\mathrm{cb}$ as the primary independence-aware $p$-value.}
  \label{tab:hemorrhage_tests}
  \scriptsize
  \setlength{\tabcolsep}{3pt}
  \begin{tabular}{llccccrc}
    \toprule
    Metric & Method & Shift [95\% CI] & $r$ & $p_\mathrm{asym}$ & $p_\mathrm{cb}$ & $q_\mathrm{BH}$ & Decision \\
    \midrule
    \multirow{8}{*}{SSIM $\uparrow$}
    & Cubic baseline     & $-0.023\ [-0.042,-0.005]$ & $-0.18$ & $2.7\!\times\!10^{-6}$  & $0.018$  & $5.4\!\times\!10^{-6}$ & $\checkmark$ \\
    & Mean baseline      & $-0.018\ [-0.037,+0.000]$ & $-0.14$ & $1.5\!\times\!10^{-4}$  & $0.056$  & $2.6\!\times\!10^{-4}$ & $\dagger$ \\
    & RIFE baseline      & $-0.029\ [-0.045,-0.013]$ & $-0.23$ & $1.4\!\times\!10^{-9}$  & $<0.001$ & $3.6\!\times\!10^{-9}$ & $\checkmark$ \\
    & FILM baseline      & $-0.030\ [-0.046,-0.014]$ & $-0.23$ & $1.2\!\times\!10^{-9}$  & $<0.001$ & $3.5\!\times\!10^{-9}$ & $\checkmark$ \\
    & SSIM$_{\mathrm{s}}$ & $-0.025\ [-0.041,-0.010]$ & $-0.25$ & $8.4\!\times\!10^{-11}$ & $0.001$  & $2.9\!\times\!10^{-10}$ & $\checkmark$ \\
    & L1                  & $-0.026\ [-0.043,-0.010]$ & $-0.25$ & $4.9\!\times\!10^{-11}$ & $<0.001$ & $2.2\!\times\!10^{-10}$ & $\checkmark$ \\
    & MS-SSIM+L1 (ref.)   & $-0.026\ [-0.043,-0.010]$ & $-0.25$ & $5.6\!\times\!10^{-11}$ & $<0.001$ & $2.2\!\times\!10^{-10}$ & $\checkmark$ \\
    & MSE                 & $-0.023\ [-0.041,-0.006]$ & $-0.20$ & $1.3\!\times\!10^{-7}$  & $0.008$  & $2.8\!\times\!10^{-7}$ & $\checkmark$ \\
    \midrule
    \multirow{8}{*}{MAE $\downarrow$}
    & Cubic baseline     & $-0.007\ [-0.011,-0.003]$ & $-0.26$ & $3.6\!\times\!10^{-12}$ & $<0.001$ & $2.1\!\times\!10^{-11}$ & $\checkmark$ \\
    & Mean baseline      & $-0.009\ [-0.013,-0.005]$ & $-0.32$ & $3.2\!\times\!10^{-17}$ & $<0.001$ & $2.6\!\times\!10^{-16}$ & $\checkmark$ \\
    & RIFE baseline      & $-0.001\ [-0.004,+0.002]$ & $-0.03$ & $0.41$                  & $0.65$   & $0.41$ & \textemdash \\
    & FILM baseline      & $+0.001\ [-0.002,+0.004]$ & $+0.03$ & $0.36$                  & $0.61$   & $0.38$ & \textemdash \\
    & SSIM$_{\mathrm{s}}$ & $+0.002\ [-0.001,+0.004]$ & $+0.09$ & $0.013$       & $0.14$   & $0.017$ & $\dagger$ \\
    & L1                  & $+0.002\ [-0.001,+0.004]$ & $+0.10$ & $0.010$                 & $0.15$   & $0.016$ & $\dagger$ \\
    & MS-SSIM+L1 (ref.)   & $+0.002\ [-0.001,+0.004]$ & $+0.10$ & $0.011$                 & $0.17$   & $0.017$ & $\dagger$ \\
    & MSE                 & $+0.002\ [-0.001,+0.004]$ & $+0.08$ & $0.028$                 & $0.22$   & $0.035$ & $\dagger$ \\
    \midrule
    \multirow{8}{*}{PSNR $\uparrow$}
    & Cubic baseline     & $+1.66\ [+0.95,+2.37]$ & $+0.37$ & $1.2\!\times\!10^{-22}$ & $<0.001$ & $1.5\!\times\!10^{-21}$ & $\checkmark$ \\
    & Mean baseline      & $+1.87\ [+1.18,+2.58]$ & $+0.42$ & $3.4\!\times\!10^{-28}$ & $<0.001$ & $8.1\!\times\!10^{-27}$ & $\checkmark$ \\
    & RIFE baseline      & $+1.74\ [+0.55,+3.02]$ & $+0.22$ & $4.4\!\times\!10^{-9}$  & $0.004$  & $1.1\!\times\!10^{-8}$ & $\checkmark$ \\
    & FILM baseline      & $+1.26\ [+0.06,+2.52]$ & $+0.16$ & $1.8\!\times\!10^{-5}$  & $0.041$  & $3.2\!\times\!10^{-5}$ & $\checkmark$ \\
    & SSIM$_{\mathrm{s}}$ & $+0.73\ [-0.40,+1.93]$ & $+0.10$ & $0.012$            & $0.21$   & $0.017$ & $\dagger$ \\
    & L1                  & $+0.61\ [-0.57,+1.82]$ & $+0.08$ & $0.039$                 & $0.31$   & $0.045$ & $\dagger$ \\
    & MS-SSIM+L1 (ref.)   & $+0.59\ [-0.51,+1.77]$ & $+0.08$ & $0.038$                 & $0.31$   & $0.045$ & $\dagger$ \\
    & MSE                 & $+0.52\ [-0.56,+1.61]$ & $+0.07$ & $0.051$                 & $0.34$   & $0.055$ & \textemdash \\
    \bottomrule
  \end{tabular}
\end{table}

%% file: sections/appendix_ssim_k_sensitivity.tex
\section{SSIM evaluation sensitivity to the stability constant \texorpdfstring{$K_2$}{K2}}
\label{sec:appendix_ssim_k}

Training-time SSIM-based losses use $K_2 = 0.4$ as the numerical-conditioning constant (intervention~1 of \Cref{sec:losses}), whereas test-time SSIM and MS-SSIM are reported at the Wang \emph{et al.}~\cite{wang1284395} default $K_2 = 0.03$ so that the numbers are comparable to prior literature. This appendix quantifies the metric impact of that reporting choice: for each trained model in \Cref{tab:test_patient}, we recompute SSIM and MS-SSIM on the same held-out test predictions at both $K_2$ settings and report the per-model mean delta. The analysis concerns the \emph{evaluation} metric only; the trained model weights are identical between the two columns.

\Cref{tab:ssim_k_delta} reports the per-model summary. Evaluating at $K_2 = 0.4$ (matching the training-time conditioning constant) uniformly inflates mean test SSIM by $0.085$--$0.093$ and mean test MS-SSIM by $0.055$--$0.063$ across the six converged models. The direction of the shift is a direct consequence of a larger $K_2$: the denominator $\sigma_x^{2} + \sigma_y^{2} + C_2$ of \Cref{eq:ssim} gains a larger additive floor, which pushes the per-window SSIM towards $1$ in the many low-texture windows characteristic of brain CT under the $[-20\,\text{HU}, 107\,\text{HU}]$ brain window.

Despite the large absolute shift, the headline model ordering is preserved. The per-model mean SSIM rankings agree across $K_2$ values at the extremes of the table (L1 at $\mathrm{lr} = 8\!\times\!10^{-4}$ ranks first under both settings; MSE ranks last under both); the only re-ordering occurs between the three mid-ranked models, whose mean SSIM differences are below $0.002$ under $K_2 = 0.03$ and below $0.0013$ under $K_2 = 0.4$, within the per-patient bootstrap CIs reported in \Cref{tab:test_patient}.

More importantly, $K_2 = 0.03$ is the more \emph{discriminating} of the two evaluation settings. The inter-model spread in mean SSIM across the six converged models is $0.0192$ at $K_2 = 0.03$ versus $0.0122$ at $K_2 = 0.4$ (a $1.57\!\times$ compression under the larger $K_2$). Paired Wilcoxon signed-rank tests on per-patient differences between the top L1 model and the MSE model reach $p = 1.86\!\times\!10^{-9}$ under $K_2 = 0.03$ for both SSIM and MS-SSIM, whereas under $K_2 = 0.4$ the same comparison weakens to $p = 4.34\!\times\!10^{-3}$ for MS-SSIM (six orders of magnitude weaker) while remaining at $p = 1.86\!\times\!10^{-9}$ for SSIM. Reporting at $K_2 = 0.03$ therefore preserves discriminating power and is the conservative choice for model comparison.

The degenerate SSIM-family checkpoint \texttt{ssim\_lr8e-4\_1b8c15} (one of the early-stopping-completed runs whose predictions are discussed in \Cref{sec:ssim-fragility}) is included in the table for completeness; its very low SSIM values at both $K_2$ settings reflect the degenerate outputs of that run rather than a property of the $K_2$ choice.

\begin{table}[!htbp]
  \centering
  \small
  \caption{Test-set SSIM and MS-SSIM under the two $K_2$ evaluation settings. $\Delta = \mathrm{mean}(K_2\!=\!0.03) - \mathrm{mean}(K_2\!=\!0.4)$ on per-patient means ($n = 30$ test patients). Training-time interventions (including $K_2 = 0.4$ in the loss) are unchanged; the model weights evaluated are identical across columns. Bold rows are the models reported in \Cref{tab:test_patient}.}
  \label{tab:ssim_k_delta}
  \begin{tabular}{lrrrrrr}
    \toprule
    \multirow{2}{*}{\textbf{Model}} & \multicolumn{3}{c}{\textbf{SSIM}} & \multicolumn{3}{c}{\textbf{MS-SSIM}} \\
    \cmidrule(lr){2-4}\cmidrule(lr){5-7}
     & $K_2\!=\!0.03$ & $K_2\!=\!0.4$ & $\Delta$ & $K_2\!=\!0.03$ & $K_2\!=\!0.4$ & $\Delta$ \\
    \midrule
    L1 ($\mathrm{lr}=8\!\times\!10^{-4}$)            & $0.8902$ & $0.9761$ & $-0.0859$ & $0.9361$ & $0.9910$ & $-0.0549$ \\
    L1 ($\mathrm{lr}=1\!\times\!10^{-4}$)            & $0.8782$ & $0.9709$ & $-0.0928$ & $0.9264$ & $0.9894$ & $-0.0630$ \\
    MSE ($\mathrm{lr}=8\!\times\!10^{-4}$)           & $0.8710$ & $0.9639$ & $-0.0929$ & $0.9315$ & $0.9912$ & $-0.0598$ \\
    SSIM ($\mathrm{lr}=3\!\times\!10^{-3}$)          & $0.8897$ & $0.9748$ & $-0.0850$ & $0.9353$ & $0.9902$ & $-0.0549$ \\
    MS-SSIM+L1 ($\mathrm{lr}=8\!\times\!10^{-4}$, \texttt{bc1d65}) & $0.8880$ & $0.9755$ & $-0.0875$ & $0.9356$ & $0.9913$ & $-0.0557$ \\
    MS-SSIM+L1 ($\mathrm{lr}=8\!\times\!10^{-4}$, \texttt{e6d845}) & $0.8856$ & $0.9740$ & $-0.0883$ & $0.9349$ & $0.9905$ & $-0.0556$ \\
    \midrule
    SSIM ($\mathrm{lr}=8\!\times\!10^{-4}$, degenerate) & $0.0672$ & $0.1488$ & $-0.0816$ & $0.0371$ & $0.1691$ & $-0.1320$ \\
    \bottomrule
  \end{tabular}
\end{table}